\documentclass[preprint,12pt,authoryear]{elsarticle}

\usepackage{amssymb}

\usepackage{color}
\usepackage{amsmath} 
\usepackage{bbold}
\usepackage{graphicx}
\usepackage{caption}
\usepackage{subcaption}

\newcommand{\Restspace}[1]{{\underline{\sigma}}_{#1}^{\rm spatial}}

\newcommand{\lspace}[2]{\hat{#1}_{#2}}
\newcommand{\lspaceall}[2]{\overline{#1}_{#2}}
\newcommand{\sspace}[2]{\widetilde{#1}_{#2}}
\newcommand{\rsspace}[2]{\check{#1}_{#2}}

\newcommand{\dint}{{\rm \,d}}

\newcommand{\ident}{\mathbb{1}}

\renewcommand{\vec}[1]{\underline{#1}}
\newcommand{\mat}[1]{\underline{\underline{#1}}}

\definecolor{darkgreen}{rgb}{0.0, 0.5, 0.0}

\journal{Journal of Computational Physics }

\begin{document}

\begin{frontmatter}




\title{XLES Part I: Introduction to Extended Large Eddy Simulation}


\author[labelCottbus,labelFU]{Christoph Glawe}

\ead{Christoph.Glawe@protonmail.com}
\author[labelCottbus]{Heiko Schmidt}
\author[labelUSA]{Alan R. Kerstein}
\author[labelFU]{Rupert Klein}

\address[labelCottbus]{BTU Cottbus-Senftenberg, 
Siemens-Halske-Ring 14, 03046 Cottbus, Germany}

\address[labelUSA]{72 Lomitas Road, Danville, CA 94526, USA}

\address[labelFU]{FU Berlin, 
Arnimallee 6, 14195 Berlin,  Germany}

\begin{abstract}

Direct numerical simulation (DNS), mostly used in fundamental turbulence 
research, is limited to low turbulent intensities due the current and future 
computer resources.
Standard turbulence models, like RaNS (Reynolds averaged 
Navier-Stokes) and LES (Large Eddy Simulation), are applied to flows in 
engineering, but they miss small scale effects, which are frequently of 
importance, see e.g. the whole area of 
reactive flows, flows with apparent Prandtl or Schmidt 
number effects, or even wall bounded flows. 
A recent alternative to these 
standard approaches is the 
one-dimensional turbulence (ODT) model, which is limited to 1D sub-domains.
In two papers we will 
provide a generalized filter strategy, called XLES (extended LES), including a 
formal theory (part I) and one special approach in the XLES family of 
models, called ODTLES (in part II (see \cite{Glawe:2014B})).
ODTLES uses an ODT sub-grid model to describe all turbulent scales not 
represented by XLES, which leaves the larger scales to be simulated 
in 3D. This allows a turbulence modeling approach with a 3D resolution mainly
independent of the turbulent intensity. Thus ODTLES is able to compute 
highly turbulent flows in domains of moderate complexity affordably and 
including the full range of turbulent and diffusive scales.
The convergence of XLES to DNS is shown and the unconventional 
XLES advection approach is 
investigated in basic numerical tests.
In part II, highly turbulent channel and duct flow results are discussed 
and show the future potential of XLES and ODTLES.

  \end{abstract}


\end{frontmatter}



\section{Introduction}
\label{s:Intro}

Turbulence is a ubiquitous effect in the physical world, with major 
technological and
environmental impacts on human society and even human existence. Turbulence 
influenced the
density and associated gravity variations that led to the formation of 
present-day galaxies, stars, and
planets. Without turbulent mixing, planetary atmospheric phenomena such as 
clouds, storms, and
precipitation essential for life on Earth would be unimaginably different. 
Turbulence in the ocean has
innumerable effects on oceanic biota, starting with the commingling of 
phytoplankton with needed
nutrients. Due to its broad influence and baffling complexity, progress in 
fundamental and practical
understanding of turbulent mixing is exceptionally challenging, yet crucial for 
scientific advancement
encompassing a wide class of problems in earth science, astrophysics, and 
engineering. For such
problems, better understanding of turbulence interactions with buoyancy effects 
and chemical and
thermodynamic processes is equally important. Due to its huge complexity, 
progress in understanding
and prediction of turbulence is extremely challenging, yet crucial for scientific 
advancement in many
disciplines.

Analytic solutions of turbulent flows cannot be derived for real life 
applications, e.g. with complex boundary conditions.
Many meteorological and technical turbulent 
flows are investigated by numerical simulations.
However the theoretical analysis of the governing 
equations is important to derive and improve simplifying models.

There are several possible levels of turbulence modeling commonly applied. The 
simplest model is no model: Direct Numerical Simulation (DNS) resolves all 
turbulent scales.
Thus impacts of modeling and numerical errors are not influencing the 
investigated physics.
DNS is widely used in fundamental research, but limited to moderate Reynolds 
numbers $Re$ (until now $Re_{\tau}\approx 5200$ for wall-bounded flows by 
\cite{Moser:2014}) and Rayleigh numbers $Ra$ due to 
the high computational effort, while many real-world flows have $Re_{\tau} 
\gtrsim 10^6$ (following \cite{Smits2013}).

In many industrial and some research applications Reynolds-Averaged 
Navier-Stokes 
equations (RaNS equations) are solved, because highly turbulent flows (e.g. 
$Re_{\tau} > 10^6$) corresponding to 
realistic flows are computationally feasible.
RaNS describes the dynamics of time-averaged fields.
The influence of fluctuating terms is modeled. 
Various models are known (see e.g. \cite{Spalart:1992} or 
an overview by \cite{Pope:2000}).
RaNS is used for steady states, but is generally 
not useful 
for computing
time-accurate flow statistics.

In recent years, Large Eddy Simulation (LES) has increasingly been used for 
industrial applied and fundamental turbulent flows.
In LES spatially filtered equations are numerically 
solved while
the unresolved sub-grid scale (SGS) terms are modeled e.g. by an eddy 
viscosity model (see e.g. \cite{Germano:1991}).



Even in very fundamental highly turbulent flows LES need to resolve a 
wide range 
of scales including at least some portion of the inertial range of 
the turbulent cascade, which limits the achievable Reynolds numbers relative to 
the achievable Reynolds numbers in RaNS simulations. 
The
parameterization of a certain range of small scales in RaNS and LES is 
especially 
problematic for multi-physics regimes such as buoyant and reacting flows because 
much of the 
complexity 
is thus relegated
to the unresolved small scales.




For further details on LES we refer to 
standard literature (e.g. by \cite{Sagaut:2006}). 
For an overview of turbulence 
properties and other model approaches we 
refer to the work by \cite{Pope:2000}.

Alternative model approaches e.g. the one-dimensional turbulence (ODT) 
model (see e.g. \cite{AR-Kerstein1999} and \cite{AR-Kerstein2001}) describe the 
3D turbulence within a 1D sub-domain including the full turbulent 
cascade, whereby the 
numerical representation of molecular diffusive effects becomes computationally 
feasible also in highly turbulent flows. Recently \cite{Meiselbach2015} 
described wall bounded flows with $Re_{\tau} \leq 6 \times 10^5$ using an 
adaptive ODT version (by \cite{Lignell2012}), which is clearly in the range 
of real-world applications, but limited to applications that are reasonable 
within a 1D 
sub-domain.  

To benefit from the ability of 1D models to describe highly turbulent flows,
several approaches combine 1D models with LES, e.g. 
LES-ODT (e.g. by \cite{Cao:2008}), $\rm 
LES/LEM$ (by \cite{Menon:2011}), and LEM3D 
(e.g. by \cite{Sannan:2013}). 
Since ODT fully resolves e.g. molecular diffusion effects, it is fairly 
non-trivial to include  it into an LES-like filter approach. 

In this work we introduce an extended LES (XLES) filter approach, which is 
tailored to include 1D models as sub-grid models and can describe highly 
turbulent flows in a domain of moderate complexity, relevant e.g. in atmospheric 
science.
To achieve this, XLES solves 2D filtered equations on a structured grid, 
maintaining one highly resolved 
Cartesian direction (this e.g. allows resolved molecular diffusion within 
an ODT sub-grid model).
To derive a preferably general model, all Cartesian directions 
are treated equally: Three 2D filters, each corresponding to one highly resolved
Cartesian direction, are applied independently to the governing equation. This 
leads to 
three coupled sets of 2D filtered equations, derived in section \ref{s:XLES}. 

Thereby XLES only needs to resolve global 3D structures
(e.g. the computational domain) independent of the turbulent cascade, while 
e.g. a one-dimensional turbulence model oriented in the respective highly 
resolved directions efficiently exploits the special symmetry of a 2D filter 
and represents the XLES microscale terms, corresponding to turbulent velocity 
scales not resolved by XLES. 
This ODTLES approach, investigated in detail in part II (\cite{Glawe:2014B}), is 
beneficial for flows with turbulent small 
scale effects playing an important role, e.g. in the combustion area, in 
buoyant stratified flows and wall-bounded flows with high turbulent intensity 
e.g. with high Reynolds or Rayleigh numbers.



This XLES based ODTLES model is an extended version of the ODTLES model, 
introduced and examined by 
\cite{RC-Schmidt2010}, \cite{ED-Gonzalez-Juez2011}, and \cite{Glawe2013}.

In this work we distinguish between the expressions XLES to 
describe the XLES approach including an approximation or model for arising 
microscale terms, 
ODTLES if these terms are in particular described by ODT and XLES-U (XLES 
unclosed), if microscale terms are neglected.

In this work the XLES approach is introduced (in section \ref{s:XLES}) and 
XLES-U is verified by numerical 
studies and applied to a turbulent channel flow (in section 
\ref{s:Properties}), followed by conclusions in section \ref{s:Conclusions}.
In Part II (see \cite{Glawe:2014B}) ODTLES is derived as one 
special approach in the XLES family of models and highly turbulent ODTLES 
channel and duct flow results are presented.


Note that in this work no Einstein summation convention is used.


\section{Extended Large Eddy Simulation (XLES)}
\label{s:XLES}

We consider the turbulent flow to be described by the incompressible 
Navier-Stokes equations for a Newtonian fluid including the conservation of 
mass in Eq. 
(\ref{eqn:EulerMass}) and momentum in Eq. (\ref{eqn:EulerMom}):
\begin{align}
\label{eqn:EulerMass}
    0 &= \sum_{j=1}^3 \partial_{x_j} u_{j}  \\ 
\label{eqn:EulerMom}
    0 &= \frac{1}{\rho_0} \partial_{x_i} p + \left[ \partial_{t}  - \nu 
\sum_{j=1}^3 \partial_{x_j}^2 \right] u_{i}  +  \sum_{j=1}^3
\partial_{x_j}  u_{j} \cdot  u_{i} 
\end{align}
with the velocity $u_i$ in Cartesian $x_i$-direction ($i=\{1,2,3 \}$), a 
constant kinematic viscosity $\nu$, and the pressure gradient $\partial_{x_i} 
p$.
For simplicity we assume in this work the constant density to be 
$\rho_0=1$.

Variables e.g. the velocities $u_i$ and operators are assumed to be described 
within a continuum. Discrete variables are marked with the superscript ${d}$ 
(e.g. discrete velocities: $u_i^{{d}}$).

Similar to LES, in XLES the velocity field is filtered. 
3D filter functions commonly applied in LES are defined as tensor 
products of 1D filter 
functions $[l_k]$  (we are using an 
operator notation) in $x_k$-direction ($k=\{1,2,3\}$; a filter 
operating on a continuum corresponds to a convolution).
To derive macroscale (to be simulated) and microscale (to be modeled) terms the 
spatial scales are separated 
for modeling purpose using these 1D filter operators:
\begin{align}
\label{eqn:GeneralScaleSepAnsatz}
       u_i
       =& \left[ 
     \ident \,\ident \,\ident
    \right]u_i  \\
    =&\left[ 
      (l_1 + (\ident -l_1) )(l_2 + (\ident -l_2))(l_3 + (\ident -l_3))
    \right]
      u_i   \nonumber \\
    \equiv&\left[ 
      (l_1 + s_1)(l_2 + s_2)(l_3 + s_3)
    \right]
      u_i \nonumber\\ 
    =& \left[  {  l_1 l_2 l_3 +s_1 l_2 l_3 + l_1 s_2 l_3 + 
l_1 l_2 s_3} + s_1 s_2 l_3 + s_1 l_2 s_3 + l_1 s_2 s_3 + s_1 s_2 s_3 \right] 
u_i  \nonumber
\end{align}  
with the unity operator $[\ident]$ and the 1D small scale operator
\begin{align}
\label{eqn:SmallScaleOp}
  [s_k] = [\ident - l_k] .
\end{align}
The tensor product ansatz causes combinations of 1D operators to be 
commutable, e.g. $s_1 l_2 l_3 = l_3 s_1 l_2$. 

In general the occurring spatial 3D scales (terms in the last row of 
Eq. (\ref{eqn:GeneralScaleSepAnsatz})) are decomposed into those to 
be simulated (resolved) and those to be modeled (unresolved). We will 
refer to this as `filter separation', because the commonly used term `scale 
separation' 
implies the simulated scales to be large.
In LES the resolved velocities correspond to the 3D large scale field $[l_1 
l_2 l_3]u_i$. Thus the LES filter separation corresponds to a 3D scale 
separation.

In table \ref{tab:CompareModels} the filter separations are compared between 
LES, 
DNS and XLES.

\begin{table}[ht]
\centering
\caption{Comparison of filter separation ansatzes}
\label{tab:CompareModels}
  \begin{tabular}{|l || l | l |}
  \hline
 Model   	 & resolved scales & unresolved scales (SGS)   \\ 
\hline \hline
  LES     		 & $\lspaceall{u}{i}^{\rm LES} =[l_1 l_2 l_3] u_i $ & 
$\sspace{u}{i}^{\rm LES}=[s_1 l_2 l_3 + l_1 s_2 l_3 + 
l_1 l_2 s_3 $\\ & & $+ s_1 s_2 l_3 + s_1 l_2 s_3 + l_1 s_2 s_3 + s_1 s_2 
s_3]u_i$ \\ \hline
  DNS     		 & $[\ident\,\ident\,\ident]u_i$ & $0$ 
\\ \hline
  XLES     		 & $\lspaceall{u}{i} = $ & $\sspace{u}{i}=$ \\ 
         		 & $[l_1 l_2 l_3+s_1 l_2 l_3 
+ l_1 s_2 l_3 +l_1 l_2 s_3]u_i$ & $[s_1 s_2 l_3 + s_1 l_2 s_3 + 
l_1 s_2 s_3 + s_1 s_2 s_3]u_i$ \\ \hline
\end{tabular}
\end{table}
The LES resolved (large scale) velocities are $\bar{u}_i^{\rm LES}= [ l_1 
l_2 l_3] u_i $, and the unresolved (SGS) velocities $\tilde{u}_i^{\rm LES}= 
[\ident-
l_1 l_2 l_3] u_i$. 
The acronym ${\rm LES}$ is used to differentiate between standard 
LES velocities and those defined in XLES.

In DNS all available scales are resolved numerically.

In XLES the resolved scales are connected to 2D filtered fields. Without loss 
of generality (w.l.o.g.) we apply $[l_1 l_2]$ leading to: 
\begin{align}
  [l_1 l_2] u_i = [l_1 l_2 l_3 + l_1 l_2 s_3 ] u_i .
\end{align}
To treat all Cartesian directions equally, all possible 2D-filtered terms $[l_2 
l_3]u_i$, 
$[l_1 
l_3]u_i$, and $[l_1 l_2]u_i$ are resolved numerically.

This XLES macroscale (to be simulated) approach can be interpreted as the LES 
macroscale ($[l_1 l_2 l_3]$) with 
additionally 1D resolved small scale (RSS) terms in all Cartesian directions 
($[s_1 l_2 l_3]+[l_1 s_2 l_3]+[l_1 l_2 s_3]$). Both, 
the XLES resolved velocity scales (investigated in this work) and the unresolved 
scales (investigated in part II (\cite{Glawe:2014B})) impose special 
requirements on the numerical scheme and modeling; techniques known from LES 
cannot be applied one-to-one to XLES.


\subsection{XLES: Spatial Filtering}
\label{ss:XLES_scales}

In XLES, 2D filter operators -each corresponding to one highly resolved 
Cartesian direction- are applied to the 
governing equations, comparable to the 3D filtering in LES.
To derive a filter approach, independent of the chosen Cartesian direction, 
three 2D filters, corresponding to three Cartesian directions, are 
applied. 
We will use a vector notation (called XLES vector 
notation), indicated by an underbar (e.g. $\vec{u}_i$). Each 
vector element contains one of the three 2D filtered velocity fields:
\begin{align}
  \begin{pmatrix}
   [l_2 l_3]{u}_{i} \\
   [l_1 l_3]{u}_{i} \\
   [l_1 l_2]{u}_{i}
 \end{pmatrix}
  = 
  \begin{pmatrix}
	l_2 l_3 & 0 & 0 \\ 
	0 & l_1 l_3 & 0 \\ 
	0 & 0 & l_1 l_2 
    \end{pmatrix}
    \begin{pmatrix}
	{u}_{i} \\
	{u}_{i} \\
	{u}_{i}
   \end{pmatrix}
  \equiv
     \mat{l}^{2D} \, \vec{u}_i
  \equiv   
     \lspace{\vec{u}}{i}
\end{align}
with the 2D filter matrix  $\mat{l}^{2D}$.

\begin{figure}
        \centering
        \begin{subfigure}[b]{0.26\textwidth}
                \includegraphics[width=\textwidth]{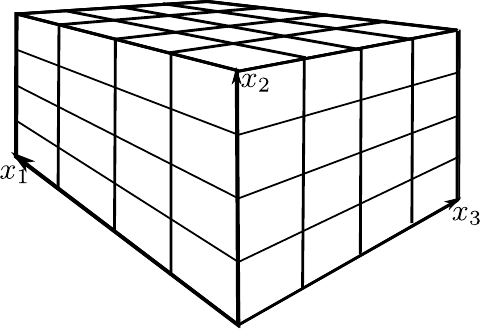}
                \caption{LES grid: \\ 
                $\lspaceall{u}{i}^{d, \textrm{LES}}\approx[l_1 l_2 l_3]u_i$.}
                \label{fig:gridLES}
        \end{subfigure}        
        \begin{subfigure}[b]{0.23\textwidth}
                \includegraphics[width=\textwidth]{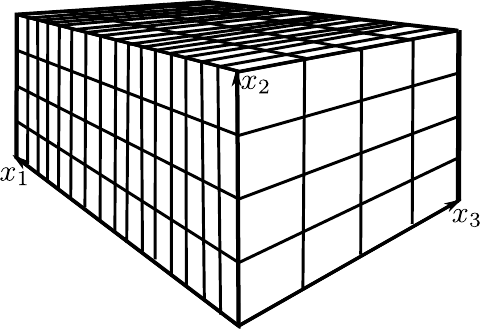}
                \caption{Grid 1:  \\ $\lspace{u}{1,i}^d\approx [l_2 l_3]u_i$.}
                \label{fig:grid1}
        \end{subfigure}
        \begin{subfigure}[b]{0.23\textwidth}
                \includegraphics[width=\textwidth]{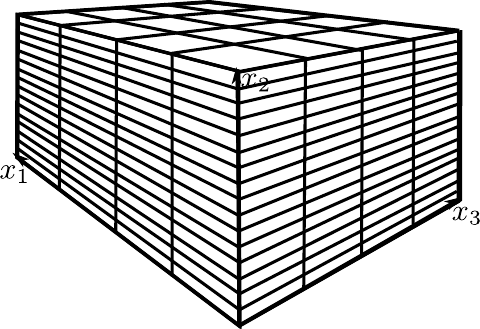}
                \caption{Grid 2:  \\ $\lspace{u}{2,i}^d\approx[l_1 l_3]u_i$.}
                \label{fig:grid2}
        \end{subfigure}
	\begin{subfigure}[b]{0.23\textwidth}
                \includegraphics[width=\textwidth]{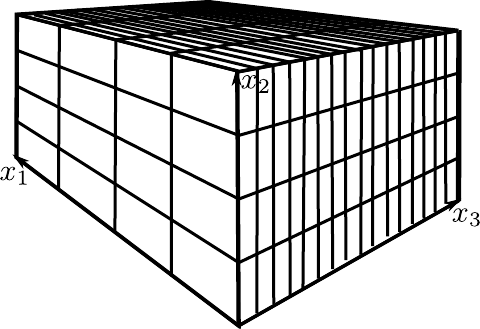}
                \caption{Grid 3:  \\ $\lspace{u}{3,i}^d\approx[l_1 l_2]u_i$.}
                \label{fig:grid3}
        \end{subfigure}
        \caption{In XLES the velocities are resolved using multiple 
XLES-grids illustrated in \ref{fig:grid1}-\ref{fig:grid3}. 3D large scale 
properties, corresponding to a standard LES grid are for illustration
represented with 
$N_{\rm LES}=4$ cells per direction in \ref{fig:gridLES}. In XLES the 3D 
large scale velocities are derived by 1D filtering the XLES 
properties: $\lspaceall{u}{i}^{\rm LES}=[l_k]\lspace{u}{k,i},\;k=\{1,2,3\}$. 
The discrete
XLES resolved small scale (`RSS') properties are represented for illustration 
by $N_{\rm RSS}=16$  cells in 
\ref{fig:grid1}-\ref{fig:grid3}. }\label{fig:grids}
\end{figure}

The 2D filtered velocity fields $\lspace{\vec{u}}{i}$ (we refer to 
$\lspace{\vec{u}}{i}$ as the XLES vector) are 
discretized using three overlapping staggered XLES-grids with face-centered 
velocities and cell-centered pressure, 
illustrated in figure \ref{fig:grid1}--\ref{fig:grid3}, where each XLES-grid 
$k=\{1,2,3\}$ discretizes one XLES vector element $\lspace{{u}}{k,i}^d$. 
The size of the discrete 1D filter $[l_k^d]$ corresponds to the large scale 
cell size $\Delta x_k^{\rm LES}$ (see figure \ref{fig:gridLES}), where the 
discrete 1D filter definition $[l_k^d]$ corresponds to a box filter 
  \begin{align}
  \label{eqn:BoxFilter}
    [l_k^d]u_i = \frac{1}{\Delta x_k^{\rm 
LES}} \int_{-\frac{\Delta x_k^{\rm 
LES}}{2}}^{\frac{\Delta 
  x_k^{\rm LES}}{2}}  
  u_i \dint x_k'.
  \end{align} 
Here the discrete mesh corresponds to the discrete filter size which is 
often called implicit filtering.

The XLES filter separation decomposes the full velocity field into three parts 
(using the XLES vector notation):
\begin{align}
 \nonumber
   \vec{u}_i 
  & \equiv
  \begin{pmatrix}
   {u}_{i} \\
   {u}_{i} \\
   {u}_{i}
 \end{pmatrix}
   = 
     \begin{pmatrix}
   [l_2 l_3] {u}_{i} \\
   [l_1 l_3] {u}_{i} \\
   [l_1 l_2] {u}_{i}
 \end{pmatrix} + 
 \begin{pmatrix}
   [ l_1 s_2 l_3 +l_1 l_2 s_3 ]u_i \\
   [ s_1 l_2 l_3 +l_1 l_2 s_3 ]u_i \\
   [ s_1 l_2 l_3 +l_1 s_2 l_3 ]u_i
 \end{pmatrix}
 + \begin{pmatrix}
   [ \mathcal{S}]u_i \\
   [ \mathcal{S}]u_i \\
   [ \mathcal{S}]u_i
 \end{pmatrix} \\
   \label{eqn_XLES_AllScalesDef}
  &\equiv  \mat{l}^{2D}\, \vec{u}_i +  \mat{C} \,\mat{s}^{1D} \mat{l}^{2D}
 \vec{u}_i + [ 
\mathcal{S}]\vec{u}_i:
\end{align}
\begin{enumerate}
\item `Directly Resolved': 
 
  The 2D filter $\mat{l}^{2D}$ applied to the full 
  velocity field leads to the XLES vector: $\lspace{\vec{u}}{i} = 
  \mat{l}^{2D}\, \vec{u}_i$. Each XLES-grid represents its own directly 
resolved velocity field distinct from the other XLES-grids.
\item `Indirectly Resolved':
 
  There are indirectly resolved  small 
scale terms (`directly resolved' by another XLES-grid): $\mat{C} 
\,\mat{s}^{1D} \mat{l}^{2D} 
   \vec{u}_i$. 
  Especially within non-linear advection terms (in Eq. 
(\ref{eqn:EulerMom})), they determine the coupling between the XLES-grids.
  The coupling matrix $\mat{C}$ and the small scale matrix $\mat{s}^{1D}$ in 
Eq. (\ref{eqn_XLES_AllScalesDef}) are:
  \begin{align}
  \mat{C}    = 
      \begin{pmatrix}
	0 	& \ident	& \ident \\
	\ident & 0	 	& \ident \\
	\ident & \ident 	& 0
    \end{pmatrix}
  \, {\rm and } \quad 
    \mat{s}^{1D} =
      \begin{pmatrix}
	s_1 		& 0		 &	0 \\
    0			& s_2	 	 &	0 \\
      0			& 0		 & 	s_3
    \end{pmatrix}.
  \end{align}
  The matrix $\mat{s}^{1D}$ defines the resolved small 
scale (RSS) velocities $\rsspace{\vec{u}}{i}$ (these are part of the XLES 
macroscale, contrary to LES):
  \begin{align}
   \label{eqn:Mom_ResolvedSmallScales}
 \mat{s}^{1D} \, \vec{u}_i   \equiv \rsspace{\vec{u}}{i}
  = \lspace{\vec{u}}{i} - \mat{l}^{1D} \lspace{\vec{u}}{i}
  = \lspace{\vec{u}}{i} - \lspaceall{\vec{u}}{i}^{\rm LES}
  \end{align}
  with
  the 1D filter matrix
  \begin{align}
  \label{eqn:XLES_1DFilter}
  \mat{l}^{1D}=
  \begin{pmatrix}
    l_1 & 0 & 0 \\ 
    0 & l_2 & 0 \\ 
    0 & 0 & l_3 
  \end{pmatrix}.
  \end{align}
  In index notation: The term (w.l.o.g.) $[l_1 s_2 l_3]u_i =[l_1 l_3]u_i - 
[l_1 l_2  l_3]u_i$  can be interpreted numerically, because   
  $\lspace{u}{2,i}=[l_1 l_3]u_{i}$ is exclusively available in 
XLES-grid $2$ and the 1D filtered XLES velocity field 
$[l_2]\lspace{u}{2,i}=[l_1 l_2 l_3]u_i$, 
corresponding to the 
LES velocity field $\lspaceall{{u}}{i}^{LES}$ (see figure \ref{fig:grids}), 
is also available in XLES-grid $2$. 


\item  `Not Resolved':

  The velocity scales $[\mathcal{S}]u_i = [s_1 s_2 l_3 + s_1 l_2 s_3 
+ l_1 s_2 s_3   + s_1 s_2 s_3]u_i$ are not resolved in any XLES-grid 
(XLES sub-grid scale (SGS) or XLES microscale). In part II 
(\cite{Glawe:2014B}) advection terms containing unresolved velocity scales are 
interpreted by the ODT model. 
\end{enumerate}

We summarize all resolved velocity scales (directly and indirectly):
\begin{align}
\label{eqn:XLES_AllResolvedScales}
 \lspaceall{\vec{u}}{i}=[l_1 l_2 l_3 +s_1 l_2 l_3 + l_1 s_2 l_3 + l_1 l_2 s_3] 
\vec{u}_i = (\mat{l}^{2D} +  \mat{C} \,\mat{s}^{1D} \mat{l}^{2D} ) \vec{u}_i.
\end{align}
A possible interpretation of the XLES macroscale is the numerical approximation 
of 
$\lspaceall{u}{i}$ instead of $\lspaceall{u}{i}^{\rm LES}=[l_1 l_2 l_3]u_i$ in 
LES (compare to table \ref{tab:CompareModels}). 
\ref{app:XLES_MatrixDerivation} shows that expressing 
$\lspaceall{\vec{u}}{i}$ in index notation automatically leads to 
three coupled XLES-grids (corresponding to Eq. 
(\ref{eqn:XLES_AllResolvedScales})).

\subsection{Momentum Conservation}
\label{ss:XLES_momentumConservation}

Similar to LES, in XLES filtered equations are solved. Contrary to LES the 
XLES equations are derived by applying a 2D filter 
matrix $\mat{l}^{2D}$ to the momentum equation Eq. (\ref{eqn:EulerMom}) 
leading to (in XLES vector notation):
\begin{align}
 \label{eqn:XLES_PreFilMomentumGridN2}
  0 =& \partial_{x_i}  \vec{\lspace{p}{}} +  \left( \partial_t  - \nu 
\sum_{j=1}^3 
\partial_{x_j}^2   \right) \vec{\lspace{u}{}}_{i} + \sum_{j=1}^{3}
 \partial_{x_j}  \vec{\lspace{u}{}}_{j}* \vec{\lspace{u}{}}_{i}
  + \sum_{j=1}^{3} \vec{\tau}_{ij}^{\rm XLES} 
\end{align}
with the 2D filtered pressure 
\begin{align}
 \vec{\lspace{p}{}} 
 =  \mat{l}^{2D}  \vec{p} =  \begin{pmatrix}
			      [l_2 l_3]p & [l_1 l_3]p &[l_1 l_2] p
			     \end{pmatrix}^T ,
\end{align}
the XLES residual stress tensors $\vec{\tau}_{ij}^{\rm XLES}$,
and the Hadamard operator $*$, an entry-wise multiplication between 
XLES vectors and matrices.

The three 2D filtered momentum equations, each describing 
three velocity components (hence the number of equations is tripled), coexist 
and are solved simultaneously.

Compared to LES advection terms $\partial_{x_j}  
\lspaceall{u}{j}^{\rm LES} \lspaceall{u}{j}^{\rm LES}$
the XLES advection terms $ \partial_{x_j} 
\vec{\lspace{u}{}}_{j}*\vec{\lspace{u}{}}_{i} =  \partial_{x_j} 
(\lspaceall{\vec{u}}{j}^{\rm LES} + 
\rsspace{\vec{u}}{j})*(\lspaceall{\vec{u}}{i}^{\rm LES} + 
\rsspace{\vec{u}}{i})$ contain additional XLES-grid specific resolved small 
scale (RSS) terms.

The contribution associated with the XLES residual stress tensors
\begin{align}
  \vec{\tau}_{ij}^{\rm XLES} = \mat{l}^{2D} \left( 
\partial_{x_j} 
\vec{{u}{}}_{j}* \vec{{u}{}}_{i}\right) 
- \partial_{x_j}  \vec{\lspace{u}{}}_{j}* \vec{\lspace{u}{}}_{i}
\end{align}
is captured through the use of some form of modeling or 
approximation.
To investigate the residual stresses in more detail, a 2D
decomposition (a modified version of  
the 3D decomposition by 
\cite{Leonard:1975}) is performed, leading to:
\begin{align}
\label{eqn:XLES_DecomposeByLeonard}
 \vec{\tau}_{ij}^{\rm XLES} 
 &= \mathcal{\vec{X}}_{ij}^{\rm XLES} +
 \mathcal{\vec{L}}^{2D}_{ij} +  
\mathcal{\vec{C}}_{ij}^{\rm XLES} +  
\mathcal{\vec{R}}_{ij}^{\rm XLES}
\end{align}
where:
\begin{itemize}
 \item XLES coupling tensor terms
 
\begin{align}
\label{eqn:XLES_Xstress}
  \mathcal{\vec{X}}_{ij}^{\rm XLES} =
  \partial_{x_j}
 \left( 
    \lspace{\vec{u}}{j} *  \mat{C}\,\mat{s}^{1D}\mat{l}^{2D}{\vec{u}}_{i} 
  + \mat{C}\,\mat{s}^{1D}\mat{l}^{2D} {\vec{u}}_{j} *  \lspace{\vec{u}}{i}
  + \mat{C}\,\mat{s}^{1D}\mat{l}^{2D}{\vec{u}}_{j} *  
\mat{C}\,\mat{s}^{1D}\mat{l}^{2D}{\vec{u}}_{i} 
\right)
\end{align}
involve `indirectly resolved' terms. 
These stress terms couple the momentum equations represented by different 
XLES-grids. A possible approximation of $\mathcal{\vec{X}}_{ij}^{\rm XLES}$ is 
investigated in
section \ref{ss:XLES_LES_Coupling}.
In LES these terms are not simulated and therefore typically 
modeled or approximated and contribute to 3D cross-stress and 3D SGS 
Reynolds stress terms.
\item 2D Leonard stresses

\begin{align}
 \label{eqn:LeonardStressdirectly}
  \mathcal{\vec{L}}^{2D}_{ij} = (\mat{l}^{2D} \partial_{x_j}
 \left( 
    \lspace{\vec{u}}{j} *  \lspace{\vec{u}}{i}
\right) 
- \partial_{x_j}  \vec{\lspace{u}{}}_{j}* \vec{\lspace{u}{}}_{i}) + (
\mat{l}^{2D} \mathcal{\vec{X}}_{ij}^{\rm XLES}
- \mathcal{\vec{X}}_{ij}^{\rm XLES})
\end{align}
describe the influence of the 2D filter on the (`directly' and 
`indirectly') resolved XLES advection terms (see section 
\ref{ss:XLES_SGS_Coupling} for details). 

\item XLES cross-stress terms

\begin{align}
\label{eqn:XLES_CrossStressTerm}
  \mathcal{\vec{C}}_{ij}^{\rm XLES} = \mat{l}^{2D} \partial_{x_j}
 \left( 
    \sspace{\vec{u}}{j} *  \lspace{\vec{u}}{i} 
  + \lspace{\vec{u}}{j} *  \sspace{\vec{u}}{i} 
\right) + 
 \mat{l}^{2D} \partial_{x_j}
 \left( 
    \sspace{\vec{u}}{j} *  \mat{C}\,\mat{s}^{1D}\mat{l}^{2D}{\vec{u}}_{i} 
  + \mat{C}\,\mat{s}^{1D}\mat{l}^{2D}{\vec{u}}_{j} *  \sspace{\vec{u}}{i} 
\right)
\end{align}
describe the interaction of resolved (`directly' and 
`indirectly') and unresolved 
XLES terms (see section \ref{ss:XLES_SGS_Coupling} for details).

\item 2D SGS Reynolds stresses

\begin{align}
  \mathcal{\vec{R}}_{ij}^{\rm XLES} = \mat{l}^{2D} \partial_{x_j}
 \left( 
    \sspace{\vec{u}}{j} *  \sspace{\vec{u}}{i}
\right)
\end{align}
describe the interaction of terms not resolved in XLES (see 
section \ref{ss:XLES_SGS_Coupling} for details).
\end{itemize}

The XLES stress terms $\mathcal{\vec{L}}^{2D}_{ij}$, 
$\mathcal{\vec{C}}_{ij}^{\rm XLES}$, 
and $\mathcal{\vec{R}}_{ij}^{\rm XLES}$ are investigated in more detail in 
section \ref{ss:XLES_SGS_Coupling} and can be interpreted in the context of the 
ODT model, as 
shown 
in part II (see \cite{Glawe:2014B}).

A 2D decomposition following the idea of \cite{Germano:1986} is also possible 
within the XLES framework.
Nevertheless the 2D decomposition following \cite{Leonard:1975} is sufficient 
to 
introduce ODT into the XLES framework and therefore used here in preparation 
for part II (\cite{Glawe:2014B}).


\subsubsection{XLES: Resolved Advection Terms}
\label{ss:XLES_LES_Coupling}

The coupling stress terms $\mathcal{\vec{X}}_{ij}^{\rm 
XLES}$ in Eq. (\ref{eqn:XLES_Xstress}) are decomposed into two terms (using 
$\rsspace{\vec{u}}{i}$ from  Eq. (\ref{eqn:Mom_ResolvedSmallScales})):
\begin{align}
\label{eqn:XLES_XstressLin}
  \mathcal{\vec{X}}_{ij}^{\rm XLES} =&   
\left( \mat{C}
 \partial_{x_j}
 \left(
	  \lspace{\vec{u}}{j} 		* \lspace{\vec{u}}{i}
	- \lspaceall{\vec{u}}{j}^{\rm LES} 	* \lspaceall{\vec{u}}{i}^{\rm 
LES}
\right)
\right)^{T}
+&  
\partial_{x_j}
 \left(
      \mat{1}\,  \rsspace{\vec{u}}{j} * \mat{1}\, \rsspace{\vec{u}}{i} 
   -  \mat{1} ( \rsspace{\vec{u}}{j} * \rsspace{\vec{u}}{i})
 \right)
\end{align}


with the matrix of ones $\mat{1}$ and where:
\begin{enumerate}
\item The first term in Eq. (\ref{eqn:XLES_XstressLin})  
    exclusively contains advection terms with advecting and advected velocities 
represented within the same XLES-grid. This can be interpreted as a 
linearization affecting the coupling, while the individual advection terms  $ 
    \partial_{x_j} \lspace{{u}}{k,j} 
    \lspace{{u}}{k,i}$ and $ \partial_{x_j}  \lspaceall{{u}}{k,j}^{\rm LES} 
    \lspaceall{{u}}{k,i}^{\rm LES}$ in each XLES-grid 
    $k$ remain non-linear.    
    We investigate w.l.o.g. one element
    of a transposed coupling vector 
    $
    \left(
     \mat{C}\partial_{x_j}      
	      \lspace{\vec{u}}{j} * \lspace{\vec{u}}{i}
      \right)^{T}
    $ in detail:
    For XLES-grid $1$ we find e.g.
    \begin{align}
      \label{eqn:CouplingExample}
      \partial_{x_j} ( [l_1 l_3] {{u}}_{j} \, [l_1 l_3] 
    {{u}}_{i})
    \end{align}
    with the velocities $[l_1 l_3] {{u}}_{j} = 
    \lspace{u}{2,j}$ ($j=\{1,2,3\}$) only discretely represented in 
XLES-grid $2$.

    This transposed coupling vector can be rearranged:
    \begin{align}
      \label{eqn:TransposeCoupling}
      (\mat{C} \,\partial_{x_j}      
	      \lspace{\vec{u}}{j} * \lspace{\vec{u}}{i})^{T} = \mat{l}^{\dag} * 
\mat{C}\,\partial_{x_j}      
	      \lspace{\vec{u}}{j} * \lspace{\vec{u}}{i}
    \end{align}
    with the matrix
    \begin{align}
    \label{eqn:XLES_FilterDag}
    \mat{l}^{\dag} = 
    \begin{pmatrix}
    \ident & {l_1^{-1}}{l_2} & {l_1^{-1}}{l_3} \\
    {l_2^{-1}}{l_1} & \ident & {l_2^{-1}}{l_3} \\
    {l_3^{-1}}{l_1} & {l_3^{-1}}{l_2} & \ident 
    \end{pmatrix} .
    \end{align}
    Using Eq. (\ref{eqn:TransposeCoupling}) the same example in Eq. 
(\ref{eqn:CouplingExample}) leads to:
    \begin{align}
        \label{eqn:CouplingExample2}
        [l_1^{-1}] [l_2]\partial_{x_j} ( [l_1 l_3] {{u}}_{j} \, [l_1 l_3] 
{{u}}_{i})
    \end{align}
    which is computed in XLES-grid $2$ and then coupled to XLES-grid $1$.
    Eq. (\ref{eqn:XLES_FilterDag}) contains deconvolution 
    operators $[l_{k}^{-1}]$, which are realizable if 
only large scale information is present. This is the case in XLES: 
the velocities $[l_1 l_3] {{u}}_{j}$ and $[l_1 l_3] {{u}}_{i}$ in the example 
in Eq. (\ref{eqn:CouplingExample2}) are large scale in the $x_1$-direction, 
which is 
the direction of the deconvolution. 
This implies that Eq. (\ref{eqn:TransposeCoupling}) is exact for a continuum.

A discrete (numerical) approximation $\mat{l}^{\dag_d} \approx \mat{l}^{\dag}$ 
is provided by the algorithm shown in section  \ref{ss:Refinement}.

 

    Using the matrix $\mat{l}^{\dag}$ (including deconvolutions) the linearized 
coupling stress terms are:
    \begin{align}
    \label{eqn:XLESLinearizedCouplingStress}
      \mathcal{\vec{X}}_{ij}^{XLES} = 
      \mat{l}^{\dag} *  \mat{C} \partial_{x_j} 
      \left(
		\lspace{\vec{u}}{j} 		* \lspace{\vec{u}}{i}
	      - \lspaceall{\vec{u}}{j}^{\rm LES} 	* 
\lspaceall{\vec{u}}{i}^{\rm 
      LES}
      \right) .
    \end{align}

\item The second term in Eq. (\ref{eqn:XLES_XstressLin}) can be expanded as 
follows:
\begin{align}
 & 
 \partial_{x_j}
 \left(
      \mat{1}\,  \rsspace{\vec{u}}{j} * \mat{1}\, \rsspace{\vec{u}}{i} 
   -  \mat{1} ( \rsspace{\vec{u}}{j} * \rsspace{\vec{u}}{i})
 \right) \\
 =& \partial_{x_j} 
 \vec{1} 
   (
      \rsspace{u}{1,j} \rsspace{u}{2,i}
    + \rsspace{u}{1,j} \rsspace{u}{3,i} 
    + \rsspace{u}{2,j} \rsspace{u}{1,i} 
    + \rsspace{u}{2,j} \rsspace{u}{3,i} 
    + \rsspace{u}{3,j} \rsspace{u}{1,i} 
    + \rsspace{u}{3,j} \rsspace{u}{2,i}
   ).\nonumber
\end{align}
These terms contain interactions of small-scale velocities resolved in 
different XLES-grids. 
We neglect these non-linear coupling terms. 
For an intact energy cascade within the turbulent flow this assumption is 
reasonable because it implies that the velocities 
$\rsspace{u}{k,j}=\lspace{{u}}{k,j} - \lspaceall{{u}}{k,j}^{\rm LES}$ are 
smaller than $\lspace{{u}}{k,j}$.
\end{enumerate}

The deconvolution within the coupling terms Eq. 
(\ref{eqn:XLESLinearizedCouplingStress}) is fundamentally different in XLES than 
in existing 
filtered treatments that do not have resolved small scales. The reason is that 
the deconvolution in 
XLES is
not intended to construct small-scale features that are otherwise non-existent, 
but rather,
to modify a small-scale structure that already exists at the resolved small 
scales,
recognizing that this structure also has low-wavenumber content. Indeed, the 
goal in
principle is to modify appropriately that low-wavenumber content while 
preserving the
high-wavenumber content to the greatest possible extent (see the example in 
\ref{app:Adv_highSSResolution_SSProperty}).

By neglecting the non-linear coupling terms 
a XLES model error is introduced:
\begin{align}
\label{eqn:XLES_ModelErrorTerm}
 \Restspace{\rm XLES} =& 
  \partial_{x_j}
 \left(
      \mat{1} \, \rsspace{\vec{u}}{j} * \mat{1}\, \rsspace{\vec{u}}{i} 
   -  \mat{1} (\rsspace{\vec{u}}{j} * \rsspace{\vec{u}}{i})
 \right) .
\end{align}

In LES-U (unclosed LES), the LES-limit of XLES-U (see section 
\ref{s:XLES_LESlimit}), the model error 
$\Restspace{\rm XLES}$ vanishes and thus comparing a convergence study of 
XLES-U 
and LES-U in section \ref{ss:Channel}) allows estimation of $\Restspace{\rm 
XLES}$.


\subsubsection{XLES-SGS: Leonard Stress, Cross-Stress and SGS Reynolds Stress 
Terms}
\label{ss:XLES_SGS_Coupling}

In this section the (to be modeled) stress tensors containing XLES 
microscale (unresolved) 
terms
$\vec{\tau}_{ij}^{\rm XLES}-\mathcal{\vec{X}}_{ij}^{\rm XLES} = 
\mathcal{\vec{L}}_{ij}^{2D} + \mathcal{\vec{C}}_{ij}^{\rm 
XLES} + \mathcal{\vec{R}}_{ij}^{\rm XLES} $ are investigated: 
\begin{itemize}
 \item The Leonard stresses $\mathcal{\vec{L}}_{ij}^{2D} = \mat{l}^{2D} 
\partial_{x_j}
 \left( 
    \lspace{\vec{u}}{j} *  \lspace{\vec{u}}{i}
    +  \mathcal{\vec{X}}_{ij}^{\rm XLES}
\right) 
- \partial_{x_j} 
 \left( 
       \lspace{\vec{u}}{j} *  \lspace{\vec{u}}{i}
    +  \mathcal{\vec{X}}_{ij}^{\rm XLES}
\right) $
  can in principle be calculated by 
  explicit filtering of the directly and indirectly resolved XLES advection 
terms. 
  
  The 2D Leonard stresses are expected to be small within a discrete 
formulation, since 
  each 2D filter size matches the corresponding XLES-grid size, similar to 
implicit filtering in LES.

  \cite{Ferziger1999} (and authors therein) report that explicit 
  filtering the LES advection terms produces immoderate dissipation and 
  neglecting the 3D Leonard stresses improves the outcome. 

  According to these arguments we neglect $\mathcal{\vec{L}}_{ij}^{2D}$.

  Other attempts to model the 3D Leonard stresses are not transferred to the 2D 
  Leonard stresses in this work.

\item The cross-stress $\mathcal{\vec{C}}_{ij}^{\rm XLES}$ and SGS Reynolds 
stress terms $\mathcal{\vec{R}}_{ij}^{\rm XLES}$

describe interactions including the XLES unresolved scales 
$\sspace{\vec{u}}{i}=[\mathcal{S}] u_i$ (see table \ref{tab:CompareModels}). 
These terms can be modeled (e.g. by 
ODT).

We decompose these stress terms:
    \begin{align}
      \label{eqn:CrossStressAndSGSReynoldsStress}
      \mathcal{\vec{C}}_{ij}^{\rm XLES} + \mathcal{\vec{R}}_{ij}^{\rm XLES} 
      =& \partial_{x_j} (\mat{l}^{2D} + \mat{l}^{\dag}*\mat{C}\, \mat{l}^{2D}) 
      \left(
	    \sspace{\vec{u}}{j} * \rsspace{\vec{u}}{i} 
	  + \rsspace{\vec{u}}{j}* \sspace{\vec{u}}{i}
      \right) \\
      +& \partial_{x_j} (\mat{l}^{2D} \quad \,\,\,\, \quad \quad \quad \,\,\,) 
      \left(
	  \sspace{\vec{u}}{j} * \lspaceall{\vec{u}}{i}^{\rm LES} + 
\lspaceall{\vec{u}}{j}^{\rm LES} * \sspace{{\vec{u}}}{i} +  
\sspace{{\vec{u}}}{i}* \sspace{{\vec{u}}}{j}
      \right)\nonumber
    \end{align}
    into:
    \begin{enumerate}
      \item  The terms $\partial_{x_j} (\mat{l}^{2D} + \mat{l}^{\dag}*\mat{C}\, 
\mat{l}^{2D}) 
      \left(
	    \sspace{\vec{u}}{j} * \rsspace{\vec{u}}{i} 
	  + \rsspace{\vec{u}}{j}* \sspace{\vec{u}}{i}
      \right)$
     
	 depend on the resolved small scale velocities $\rsspace{\vec{u}}{j}$ 
	 exclusively available in one XLES-grid.

     \item  The terms $\partial_{x_j} \mat{l}^{2D}  
      \left(
	  \sspace{{\vec{u}}}{j}* \lspaceall{\vec{u}}{i}^{\rm LES} + 
\lspaceall{\vec{u}}{j}^{\rm     LES} *
    \sspace{{\vec{u}}}{i} +  \sspace{{\vec{u}}}{i} *\sspace{{\vec{u}}}{j}
      \right)$
     
	 are independent of the XLES-grid (equal in all XLES-grids). 
	 If a modeling approach is simultaneously applied in different 
XLES-grids, a SGS-coupling is required to guarantee a consistent 3D large scale 
field. 
Ad hoc introduction of an additional coupling, e.g. $\left(\mat{l}^{2D} + 
\mat{l}^{\dag}*\mat{C}\,\mat{l}^{2D} \right)
            \left(
	 	  \sspace{{\vec{u}}}{j}* \lspaceall{\vec{u}}{i}^{\rm LES} + 
\lspaceall{\vec{u}}{j}^{\rm     LES} *
    \sspace{{\vec{u}}}{i} +  \sspace{{\vec{u}}}{i} *\sspace{{\vec{u}}}{j}
      \right)$, would lead to double counting of small scale terms available in 
all XLES-grids. 
Nevertheless the exact relation
	  \begin{align}
	     &\mat{l}^{2D}    
	     \left(
	 	  \sspace{{\vec{u}}}{j}* \lspaceall{\vec{u}}{i}^{\rm LES} + 
\lspaceall{\vec{u}}{j}^{\rm     LES} *
    \sspace{{\vec{u}}}{i} +  \sspace{{\vec{u}}}{i} *\sspace{{\vec{u}}}{j}
      \right) \\
      =& \frac{1}{3}\left(\mat{l}^{2D} + 
\mat{l}^{\dag}*\mat{C}\,\mat{l}^{2D} \right)
            \left(
	 	  \sspace{{\vec{u}}}{j}* \lspaceall{\vec{u}}{i}^{\rm LES} + 
\lspaceall{\vec{u}}{j}^{\rm     LES} *
    \sspace{{\vec{u}}}{i} +  \sspace{{\vec{u}}}{i} *\sspace{{\vec{u}}}{j}
      \right). \nonumber
	  \end{align} 
    \end{enumerate}
avoids double counting due to the factor $\frac{1}{3}$ (proof of 
$\mat{l}^{2D}=  \frac{1}{3}\left(\mat{l}^{2D} + 
\mat{l}^{\dag}*\mat{C}\,\mat{l}^{2D} \right)$ by insertion).
    
    We reformulate and summarize the XLES microscale terms in Eq. 
(\ref{eqn:CrossStressAndSGSReynoldsStress}):
    \begin{align}
      \mathcal{\vec{C}}_{ij}^{\rm XLES} + \mathcal{\vec{R}}_{ij}^{\rm XLES} 
      = (\mat{l}^{2D} + \mat{l}^{\dag}*\mat{C}\, \mat{l}^{2D}) 
    \mathcal{\vec{M}}_{ij}
    \end{align}
    with 
    \begin{align}
     \label{eqn:XLES:SGSterms}
    \mathcal{\vec{M}}_{ij}
    =\partial_{x_j}  
    \left(
	    \sspace{\vec{u}}{j} * \rsspace{\vec{u}}{i} 
	  + \rsspace{\vec{u}}{j}* \sspace{\vec{u}}{i}
	  + \frac{1}{3} 
	  \left(
		  \sspace{\vec{u}}{j} *	\lspaceall{\vec{u}}{i}^{\rm LES}
	      +   \lspaceall{\vec{u}}{j}^{\rm LES}*\sspace{\vec{u}}{i}
	      +   \sspace{\vec{u}}{i}	*\sspace{\vec{u}}{j}
	  \right)
      \right) .
    \end{align}
    Each vector element $\mathcal{{M}}_{k,ij}$ contains the 
unresolved terms (SGS) in XLES-grid $k$ which are coupled to the other 
XLES-grids by the SGS coupling terms $\mat{l}^{\dag}*\mat{C}\,\mat{l}^{2D} 
\mathcal{\vec{M}}_{ij}$. 
\end{itemize}

Summarizing section \ref{ss:XLES_LES_Coupling} and \ref{ss:XLES_SGS_Coupling} 
we can write the XLES momentum equations:
\begin{align}
  0 =& \partial_{x_i}  \vec{\lspace{p}{}} +  \left( \partial_t  - \nu 
\sum_{j=1}^3 
\partial_{x_j}^2   \right) \vec{\lspace{u}{}}_{i} + \sum_{j=1}^{3}
 \partial_{x_j}  \vec{\lspace{u}{}}_{j}* \vec{\lspace{u}{}}_{i}
 + \Restspace{\rm XLES}	\nonumber\\
  +& \sum_{j=1}^{3} 
      \left(
	      \mat{l}^{2D}  \mathcal{\vec{M}}_{ij}
      \right)  
    + \Restspace{\rm SGS} \label{eqn:XLES_PreFilMomentumGridN} \\
  +& \sum_{j=1}^{3} 
      \left(
	    \mat{l}^{\dag} *  \mat{C} \,
	    \partial_{x_j}
		( \lspace{\vec{u}}{j} 		* \lspace{\vec{u}}{i}	      
		- \lspaceall{\vec{u}}{j}^{\rm LES} 	* 
\lspaceall{\vec{u}}{i}^{\rm LES} )
      +
	    \mat{l}^{\dag} *  \mat{C} \,\mat{l}^{2D}\,\mathcal{\vec{M}}_{ij}
      \right)  \nonumber .
\end{align}
The last line in Eq. (\ref{eqn:XLES_PreFilMomentumGridN}) corresponds to the 
full coupling, including the SGS terms and the indirectly resolved terms.

Additionally to the error term $\Restspace{XLES}$ in Eq. 
(\ref{eqn:XLES_ModelErrorTerm}) neglecting the 
Leonard stresses introduces the SGS error term:
\begin{align}
   \label{eqn:SGS_ModelErrorTerms}
   \Restspace{\rm SGS} = 
          \mathcal{\vec{L}}_{ij}^{2D}.
\end{align}
Please note at this stage no concrete SGS model is introduced to approximate 
$\mathcal{\vec{M}}_{ij}$.

\subsubsection{XLES: Mass Conservation}
\label{s:mass}

In the incompressible flow regime, the filtered velocity fields need to be 
divergence free to ensure mass conservation.

The 2D filter matrix $\mat{l}^{2D}$ is applied to the mass conservation 
equation Eq. (\ref{eqn:EulerMass}) (in the XLES vector notation): 
\begin{align}
\label{eqn:XLES_massTime}
 0 &= \sum_{i=1}^3 \partial_{x_i} \mat{l}^{2D}  {\vec{u}}_{i}
= \sum_{i=1}^3 \partial_{x_i} \lspaceall{\vec{u}}{i}^{\rm LES}  +    
\sum_{i=1}^3 
\partial_{x_i}  \rsspace{\vec{u}}{i}
\end{align}
which is decomposed into 3D large scale velocity fields 
$\lspaceall{\vec{u}}{i}^{\rm 
LES}$ and resolved small scale (RSS)
velocities $\rsspace{\vec{u}}{i}$.

A possible approach to solve Eq. (\ref{eqn:XLES_massTime}) is to ensure mass 
conservation for both decomposed velocity fields $\lspaceall{\vec{u}}{i}^{\rm 
LES}$ and $\rsspace{\vec{u}}{i} = \mat{s}^{1D} \vec{u}_i$:
\begin{enumerate}
 \item The equation $ 0= \sum_{i=1}^3 \partial_{x_i}  
      \lspaceall{\vec{u}}{i}^{\rm LES}$ 
      
      corresponds to the mass conservation in LES schemes. Standard discrete  
approaches applied in LES can be used. 
      In our implementation a pressure Poisson equation is solved. This leads 
      to a large scale pressure field
      $\lspaceall{\vec{p}}{}^{\rm LES} = \mat{l}^{3D} \vec{p}$, whose gradient 
    enforces a divergence free velocity field 
    $\lspaceall{u}{i}^{\rm LES}$ by solving $\partial_t  
\lspaceall{u}{i}^{\rm LES}   + 
    \partial_{x_i} 
    \lspaceall{\vec{p}}{}^{\rm LES} = 0$  (for details we refer to 
standard 
textbooks, e.g. \cite{Ferziger1999}).

\item The equation  $0= \sum_{i=1}^3 
      \partial_{x_i} \rsspace{\vec{u}}{i}$,

      corresponding to the RSS velocity fields, is discretely fulfilled without 
additional effort under some conditions. 

      
      In \ref{app:Mass_ResolvedSmallScales} the XLES mass conservation is 
derived in detail and three conditions are identified to ensure the RSS 
velocities to be divergence free:
      \begin{itemize}
      \item Consistency condition: $\lspaceall{{u}}{k,i}^{\rm LES}$ is 
	equal in all XLES-grids $k$ which is valid due to coupling (see section 
	\ref{s:XLES_LESlimit} for details).

       \item A divergence free 3D large scale velocity field which is valid 
after the standard pressure projection.

       \item A discrete 1D box filter defined in Eq. (\ref{eqn:BoxFilter}) is 
used.
	  Here we introduce a coarse-grained (large scale) and staggered 
control volume of the size $\Delta x_k^{\rm LES}$ (this cell size equals the  
filter size).   
	  Please note this filter definition is only used explicitly to create 
$\lspaceall{u}{i}^{\rm LES}=[l_k]\lspace{u}{k,i}$ within the coupling and the 
mass 
conservation. The 2D filter $\mat{l}^{2D}$ is not applied explicitly to any 
term.
      \end{itemize}
\end{enumerate}

In summary the discrete mass conservation is assured by a
standard 3D approach (large scale) with $ \mathcal{O}(N_{\rm LES}^3)$ cells, if 
a box filter and a staggered grid is used.


In XLES the velocity fields, including consistent 3D large scale informations,
are discretely interpreted in three XLES-grids simultaneously.
For such a system an existing redundancy of the discrete velocity fields can be 
exploited by
deriving the velocity components $\lspace{\vec{u}}{k,k}^d$ within one 3D large 
scale cell due 
to a direct solution, w.l.o.g. in XLES-grid $1$:
\begin{align}
\label{eqn:PoorMansProjection}
    \lspace{{u}}{1,1}^d\left( \frac{-\Delta x_1^{\rm LES}}{2} + x_1\right) 
    &=    \lspaceall{{u}}{1}^{d,\textrm{ LES}} \left(-\frac{\Delta 
    x_1^{\rm LES}}{2}\right)\\
      &- \int_{-\frac{\Delta x_1^{\rm LES}}{2}}^{-\frac{\Delta x_1^{\rm LES 
}}{2} + x_1 } 
	\partial_{x_{2}} \lspace{{u}}{1,{2}}^d \dint{x'_1}
      - \int_{-\frac{\Delta x_1^{\rm LES}}{2}}^{-\frac{\Delta x_1^{\rm 
LES}}{2} + x_1} 
	\partial_{x_{3}} \lspace{{u}}{1,{3}}^d \dint{x'_1} 
\nonumber 
\end{align}
for $x_1\leq \Delta x_1^{\rm LES} $. Eq. (\ref{eqn:PoorMansProjection}) is a 
semi discrete interpretation of Gauss's theorem for a divergence free 
velocity field tailored for velocities in XLES-grid $1$ (see figure 
\ref{fig:grid1}) within one 3D large scale cell (see figure \ref{fig:gridLES}) 
on a staggered face centered grid (no interpolation necessary). 

Since w.l.o.g. $\lspace{u}{1,1}$ is specified due to Eq. 
(\ref{eqn:PoorMansProjection}), the momentum equation Eq. 
(\ref{eqn:XLES_PreFilMomentumGridN}) in XLES-grid $1$ only needs to be solved 
for $i=\{2,3\}$.
In consequence  $6$ momentum equations are dynamically solved ($2$ velocity 
components in each of $3$ XLES-grids), while 3 velocity components can be 
derived by Eq. (\ref{eqn:PoorMansProjection}).

These reduced momentum equations can be expressed by multiplying a 
Kronecker delta operator matrix
\begin{align}
\label{eqb:DiracDeltaOperatorMatrix}
 \mat{\ident}-\mat{\delta_{i}}  = 
  \begin{pmatrix}
     1-\delta_{1i} & 0 & 0 \\
     0 & 1-\delta_{2i} & 0 \\
     0 & 0 & 1-\delta_{3i} 
  \end{pmatrix}
  , \, {\rm with} \,
  ( 1-\delta_{ki}) = \begin{cases}
                    0 , \, {\rm if} \, k = i \\
                    1 , \, {\rm else}
		   \end{cases}
\end{align}
giving:
\begin{align}
 \label{eqn:XLES_MomentumInclMass}
  0 =& (\mat{\ident}-\mat{\delta_{i}})  
  \left( 
    \partial_{x_i}  \vec{\lspaceall{p}{}}^{\rm LES} +  \left( \partial_t 
 - \nu 
\sum_{j=1}^3 
\partial_{x_j}^2   \right) \vec{\lspace{u}{}}_{i} + 
\sum_{j=1}^{3}
 \partial_{x_j}  \vec{\lspace{u}{}}_{j}* \vec{\lspace{u}{}}_{i}
  \right)
    \\
  +&
  (\mat{\ident}-\mat{\delta_{i}})  
  \left( 
  \sum_{j=1}^{3} 
      \left(
	    \mat{l}^{\dag} * 
	       \mat{C}  \partial_{x_j}
	       (
		  \lspace{\vec{u}}{j} 		* \lspace{\vec{u}}{i}	      
		- \lspaceall{\vec{u}}{j}^{\rm LES} 	* 
\lspaceall{\vec{u}}{i}^{\rm LES} 
		)
      \right)
    \right)  + (\mat{\ident}-\mat{\delta_{i}}) \Restspace{}\nonumber
 \\
  +& (\mat{\ident}-\mat{\delta_{i}}) \sum_{j=1}^{3} 
      \left(
	      \mat{l}^{2D}  \vec{\mathcal{M}}_{ij}
      \right)  
  + (\mat{\ident}-\mat{\delta_{i}}) \sum_{j=1}^{3} 
      \left(
	    \mat{l}^{\dag} * 
	       \mat{C}\,\mat{l}^{2D} \vec{\mathcal{M}}_{ij}
      \right) \nonumber .
\end{align}
Please note the Kronecker delta operator matrix is also applied to the 
unresolved 
terms 
$\vec{\mathcal{M}}_{ij}$ and therefore can be exploited within a model 
approach.
Additionally the factor $\frac{1}{3}$ within 
Eq. (\ref{eqn:XLES:SGSterms}) can be replaced by
$\frac{1}{2}$.

Since in XLES no 3D small scale velocity field is defined, small scale 
pressure 
effects can only be captured by the modeling approach (additionally 
to the modeling of the SGS advection terms $\vec{\mathcal{M}}_{ij}$).

\section{Properties of the XLES Approach}
\label{s:Properties}

\subsection{XLES: Deconvolution}
\label{ss:Refinement}

In sections \ref{ss:XLES_LES_Coupling} and \ref{ss:XLES_SGS_Coupling} coupling 
terms between the XLES-grids are introduced.
A discrete approximation of the transposed coupling matrix 
$(\mat{C}\,\mat{l}^{\rm 2D})^T =
\mat{l}^{\dag}* \mat{C}\,\mat{l}^{\rm 2D}$ is required to numerically 
represent 
these coupling terms.

Within the matrix $\mat{l}^{\dag}$ (see Eq. (\ref{eqn:XLES_FilterDag})) an 
discrete deconvolution operator $[l_k^{-1_d}]$ has to be interpreted.

In general a velocity field $u$ (indices are skipped in this section and all 
variables are assumed to be discrete representations) 
cannot be reconstructed exactly ($[l^{-1} l] 
\neq1$), because by filtering (e.g. $u^{\rm 
LES}=[l]u$) information get lost. This is not recoverable, unless 
only 
large scale information is present in the full spectrum, which is fortunately 
the case in XLES, as shown by an example in Eq. (\ref{eqn:CouplingExample2}).

The deconvolution within XLES is not intended to construct small-scale 
features, but rather, to modify the low-wavenumber content of an existing fully 
resolved property. 
This modification must meet several requirements:
\begin{itemize}
 \item Integral 
constraints are imposed
to satisfy conservation laws at the level of the individual 3D coarse-grained 
cells. This means the box filtered large scale field needs to be preserved by 
the 
deconvolution, e.g. to enable the consistency preservation of the 3D velocity 
field required for the mass conservation and the LES limit of the XLES model 
(in section 
\ref{s:XLES_LESlimit}):
\begin{align}
 \label{eqn:Ref_condition}
 [l][l^{-1}]=1 .
\end{align}
 \item The deconvolution must in some sense modify the large scale structure 
along the
resolved direction while in some sense preserving the small-scale structure. In 
this sense
this deconvolution is a reconstruction of existing structure rather than a 
construction of
something that does not otherwise exist. For this reason the term 
reconstruction 
has been
used to describe it (\cite{RC-Schmidt2010} and \cite{McDermott2005}).
\end{itemize}

In principle the deconvolution approach is not restricted to one particular 
filter definition (e.g. spectral filter, box filter, Gaussian filter), but the 
mass conservation is greatly simplified if a box filter 
is used, as shown in \ref{app:Mass_ResolvedSmallScales}.

In this section we introduce a discrete approximation of the deconvolution 
operator $[l^{-1}]$ for the purpose of computing a highly 
resolved velocity field $u_a \approx [l^{-1_d}]u^{\rm LES}$ ($a$ indicates an 
approximation by discretely interpreting $[l^{-1}]$) from a box filtered 
field $u^{\rm LES} = [l]u$.



%

\cite{RC-Schmidt2010} introduces a recursive and very fast algorithm 
approximating $u_a \approx [l^{-1}] u^{\rm LES}$ with 8th order accuracy 
without changing the box filtered field ($[l]u_a = u^{\rm LES} \equiv [l]u$ 
, see Eq. (\ref{eqn:Ref_condition})). For algorithmic details, we refer to 
\cite{RC-Schmidt2010}, \cite{ED-Gonzalez-Juez2011}, and \cite{McDermott2005}. 

The algorithm produces artificial local extrema in areas of monotone large scale 
fields with large 
gradients, e.g. occurring near a wall as reported by \cite{McDermott2005}.
Similar effects are known from deconvolution approaches within the fields of 
image and signal processing. 
We expand this algorithm by applying the linear slope limiter by 
\cite{Burbeau:2001} after each recursive step in the area of monotone 
velocities. Hereby monotonicity is defined locally for one cell by 
including its neighbors.




\begin{figure}
        \centering
	    \includegraphics[width=0.75\textwidth]
		{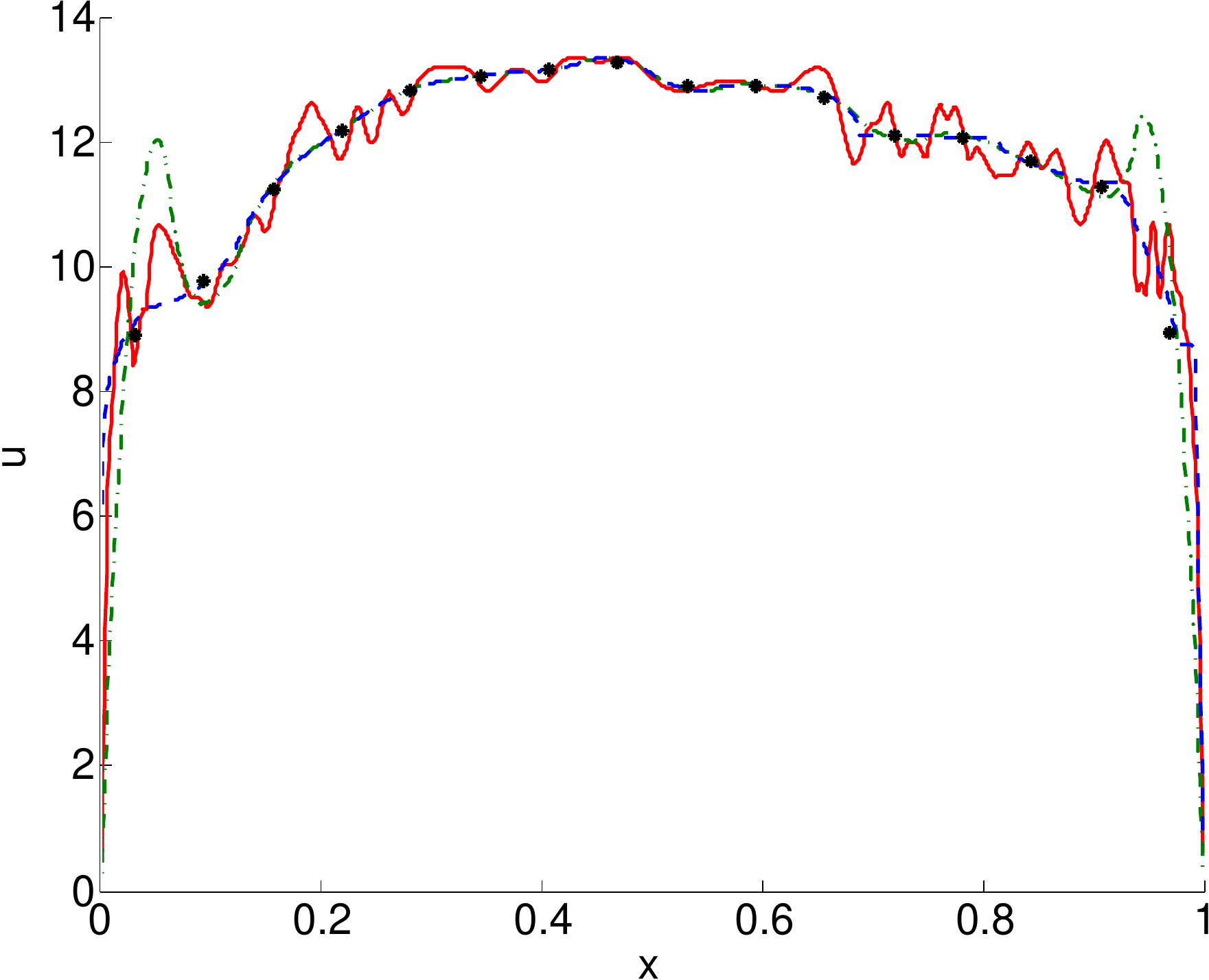}
	    \caption{A representative instantaneous streamwise velocity profile 
$u$
(\textcolor{red}{solid})  for a turbulent channel ($Re_{\tau}=1020$) with 
$N_{\rm LES}=16$ and $N_{\rm RSS}=1024$ is shown.
The large scale velocity $u^{\rm LES}=[l]u$ is represented with points.
The deconvolution $u_a \approx [l^{-1}]u^{\rm LES}$ using the original 
algorithm by 
\cite{RC-Schmidt2010} is \textcolor{darkgreen}{dashed-dotted} and the limited 
algorithm is \textcolor{blue}{dashed}. For both algorithms (limited and 
unlimited) the reconstructed field $u_a$ preserves the box-filtered values 
$[l]u_a = u^{\rm LES}$, but without the limiter unphysical overshoots occur 
near 
the walls.}
	    \label{fig:RefineVelocity}
\end{figure}

The properties of the original algorithm and the limited one are demonstrated 
for a realistic 
instantaneous streamwise velocity profile $u$ for a turbulent channel case 
with high gradients at the walls, shown in Figure \ref{fig:RefineVelocity} 
(please note in XLES only 
large scale information is deconvolved).

The numerical deconvolution error due to the algorithm described can be written 
formally:
\begin{align}
   \Restspace{a}
= ( \mat{l}^{\dag} *\mat{C}  -  \mat{l}^{\dag_a} *\mat{C} ) 
   \left(
    \partial_{x_j} 
	\left(
		  \lspace{\vec{u}}{j} 		* \lspace{\vec{u}}{i}
		- \lspaceall{\vec{u}}{j}^{\rm LES} 	* 
  \lspaceall{\vec{u}}{i}^{\rm LES}
	\right) 
	+
	\mat{l}^{2D} \mathcal{\vec{M}}_{ij}
   \right).	
\end{align}
There is no spectrally sharp way to implement this type of deconvolution. 
Various non-equivalent procedures are possible in principle. 
\cite{RC-Schmidt2010} describe a particular approach that is adopted here with a 
technical 
modification
that improves its behavior near walls. To summarize, although the inverse 
operator $l^{-1}$ arises as a natural and necessary consequence of the XLES 
ansatz, the ansatz per se
does not uniquely define its meaning nor guarantee that it can be specified in 
a 
way that
is free of unintended artifacts.


\subsection{XLES: Consistency Preservation and `LES limit'}
\label{s:XLES_LESlimit}

Both, the XLES mass conservation in section \ref{s:mass} and the coupling  
terms in section \ref{ss:XLES_LES_Coupling} and \ref{ss:XLES_SGS_Coupling} 
assume the 3D large scale velocity field to be consistent, meaning each 
XLES-grid $k$ contains the identical 3D large scale velocity field: 
$\lspaceall{u}{i}^{\rm LES} = \lspaceall{u}{k,i}^{\rm LES} \equiv [l_k] 
\lspace{u}{k,i}$.

In the XLES-vector notation this condition can be written as:
\begin{align}
 \label{eqn:XLES_ConsistentFields}
   \mat{l}^{1D} \lspace{\vec{u}}{i} = \lspaceall{\vec{u}}{i}^{\rm LES} = 
   \lspaceall{{u}}{i}^{\rm LES}     \begin{pmatrix}
                                      1 \\ 1 \\ 1
                                    \end{pmatrix}
\end{align}
with the 1D filter matrix defined in Eq. (\ref{eqn:XLES_1DFilter}). This means 
consistency preserving terms should be independent of the XLES-grid $k$.

We assume that the initial conditions are consistent (this can easily be 
achieved) and need to prove a consistency preserving XLES-advancement 
including advective and diffusive terms and the sub-grid modeling. 


The `directly resolved' advection terms themselves ($\partial_{x_j} 
\lspace{\vec{u}}{j} * \lspace{\vec{u}}{j}$) are violating the consistency 
condition, but by including the corresponding coupling (indirectly 
resolved terms): $\partial_{x_j} \lspace{\vec{u}}{j} * 
\lspace{\vec{u}}{j} + \mathcal{\vec{X}}_{ij}^{\rm XLES}$, the consistency is
preserved. The same is valid for microscale terms and their couplings:
$\mat{l}^{2D} \mathcal{\vec{M}}_{ij} + \mat{l}^{\dag_a} * 
\mat{C}\,\mat{l}^{2D} \mathcal{\vec{M}}_{ij} $. 

By rearranging the XLES advection terms and their corresponding
 couplings, we find:
\begin{align}
   \lspace{\vec{u}}{j} * \lspace{\vec{u}}{i} + 
   \mathcal{\vec{X}}_{ij}^{\rm XLES} = \lspaceall{\vec{u}}{j} * 
\lspaceall{\vec{u}}{i} 
\end{align}
with $\lspaceall{u}{k,i}= [l_1l_2l_3+s_1l_2l_3+l_1s_2l_3+l_1l_2s_3]u_i$, 
which is equal in all XLES-grids $k$. 

The same holds for the XLES sub-grid terms 
$\mat{l}^{2D}\mathcal{\vec{M}}_{ij}$ and their corresponding couplings 
$\mat{l}^{\dag_a} * \mat{C}\,\mat{l}^{2D}\mathcal{\vec{M}}_{ij}$ (proof by 
insertion).

\begin{figure}
        \centering
	    \includegraphics[width=0.5\textwidth]
		{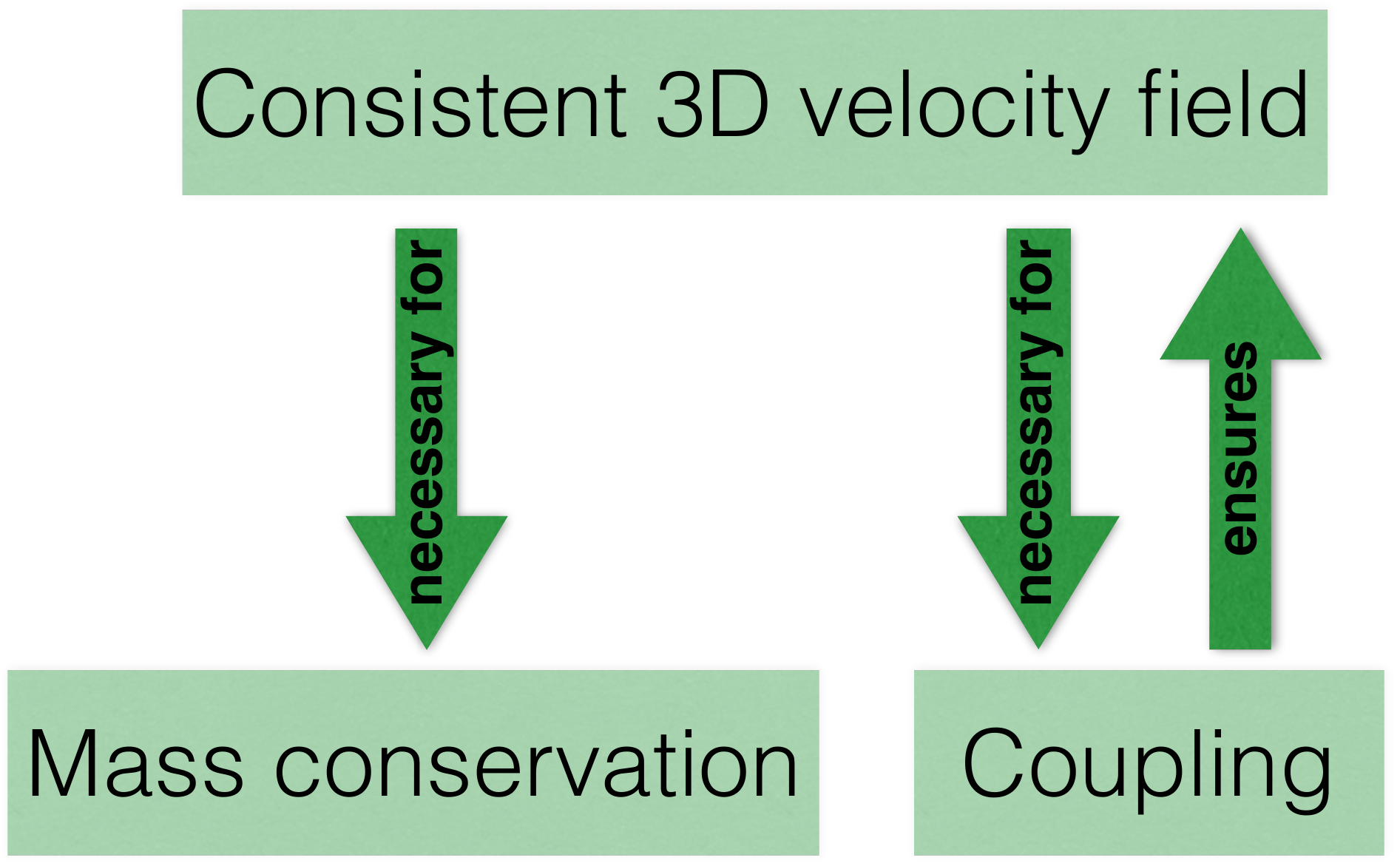}
	    \caption{On the one hand the consistency preservation is invoke 
before the advection step to define the 
coupling terms. On the other hand the coupling guarantees the consistency to be 
still preserved after the advection step.
}
	    \label{fig:ConsistentLargeScaleField}
\end{figure}

In consequence the coupling terms are essential for a consistent 3D velocity 
field within XLES, 
as illustrated in figure \ref{fig:ConsistentLargeScaleField}.

In the implementation of XLES we check that the 3D large scale velocity field 
is still consistent every $100$ timesteps to avoid numerical errors 
violating consistency preservation. All computations show the consistency to 
be limited by the floating point accuracy.

In the limit $N_{\rm RSS} \rightarrow N_{\rm LES}$ the 2D filtered XLES-U
equations collapse to the 3D filtered LES-U equations in each 
XLES-grid, because the RSS velocities vanish: $\rsspace{u}{i}= 
\lspace{u}{i} - \lspaceall{u}{i}^{\rm LES} = 0$

For $N_{\rm LES}  \rightarrow N_{\rm RSS}( \geq N_{\rm DNS})$ in each 
direction, all velocity scales are resolved by the 3D grid and XLES-U 
(and LES-U) converges to DNS. 

Within the ODT model, applied as an SGM in part II (\cite{Glawe:2014B}), all 
turbulent events are suppressed in the DNS-limit. Thus DNS is also a 
distinguished limit of ODTLES.


\subsection{XLES: Coupled Advection Scheme}
\label{ss:XLES_Numerics}


The characteristic shape of the XLES-grids (figure 
\ref{fig:grid1}--\ref{fig:grid3}) is considered when choosing the 
numerical schemes to be implemented, e.g. an implicit time 
discretization in the highly resolved direction is applied, while explicit 
time schemes are used to advect large scale properties. Therefore 
different numerical advection schemes 
are mixed due to the XLES coupling terms, because the same property, 
represented by several XLES-grids is simultaneously advanced by different
numerical schemes. 
Additionally the coupling requires a deconvolution function, which interacts 
with these numerical schemes.

In this section we present the numerical properties of this coupled advection 
scheme. For simplicity the 3D large scale velocity fields 
$\lspaceall{\vec{u}}{i}^{\rm LES}$ 
are resolved by $N_{\rm LES}$ cells in each 
direction.  The 
resolved 
small scale (RSS) properties are discretized with $N_{\rm RSS}$ cells in all 
XLES-grids.

The coupled advection scheme numerically approximates the equation
\begin{align}
 \label{eqn:XLES_CoupledAdvetionEquation}
  0 =&  (\mat{\ident} - \mat{\delta_i})
  \left(
	\partial_t  \vec{\lspace{u}{}}_{i} + \sum_{j=1}^{3}
      \partial_{x_j}  \vec{\lspace{u}{}}_{j}* \vec{\lspace{u}{}}_{i}
	+ \sum_{j=1}^{3} 
	    \left(
		  \mat{l}^{\dag} * 
		     \mat{C} \partial_{x_j}
		      ( \lspace{\vec{u}}{j} 		* \lspace{\vec{u}}{i}	
		      - \lspaceall{\vec{u}}{j}^{\rm LES} 	* 
			\lspaceall{\vec{u}}{i}^{\rm LES} )
	    \right)
  \right)
\end{align}
(according to Eq. (\ref{eqn:XLES_MomentumInclMass}); SGS terms are not 
considered).

Fundamental properties of the coupled advection scheme, e.g. the 
numerical dissipation and 
dispersion, can be demonstrated by solving a one dimensional linear advection 
problem.
Then a constant wave speed $c_j$ replaces the advecting velocities in 
Eq. (\ref{eqn:XLES_CoupledAdvetionEquation}): $  \vec{\lspace{u}{}}_{j} = 
\lspaceall{\vec{u}}{j}^{\rm LES} = c_j 
\begin{pmatrix}
  1 & 1 & 1
\end{pmatrix}^T $. 
In this section the linearized coupled advection scheme is investigated, while 
the full non-linear coupled advection scheme is studied in section 
\ref{ss:Channel}.

W.l.o.g. the advected velocity ${\vec{u}}_{1}$ is 
represented by the two staggered XLES-grids containing $\lspace{{u}}{{2,1}}$ 
and 
$\lspace{{u}}{{3},1}$ respectively while $\lspace{{u}}{1,1}$ is evaluated using 
Eq. 
(\ref{eqn:PoorMansProjection}).

\begin{figure}
        \centering
        \begin{subfigure}[b]{0.49\textwidth}
                \includegraphics[width=\textwidth]{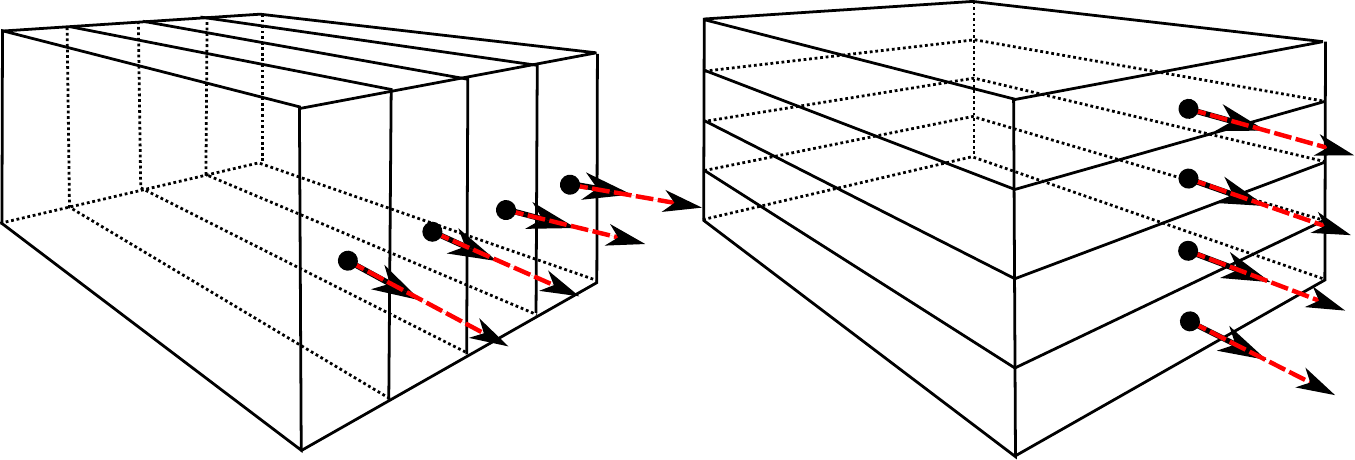}
                \caption{Case $1$: $j=i$ in 
Eq. (\ref{eqn:XLES_CoupledAdvetionEquation}). Coarsely resolved advection in 
XLES-grid $3$ (left) and XLES-grid $2$ (right).}
                \label{fig:case1}
        \end{subfigure}        
        \begin{subfigure}[b]{0.49\textwidth}
                \includegraphics[width=\textwidth]{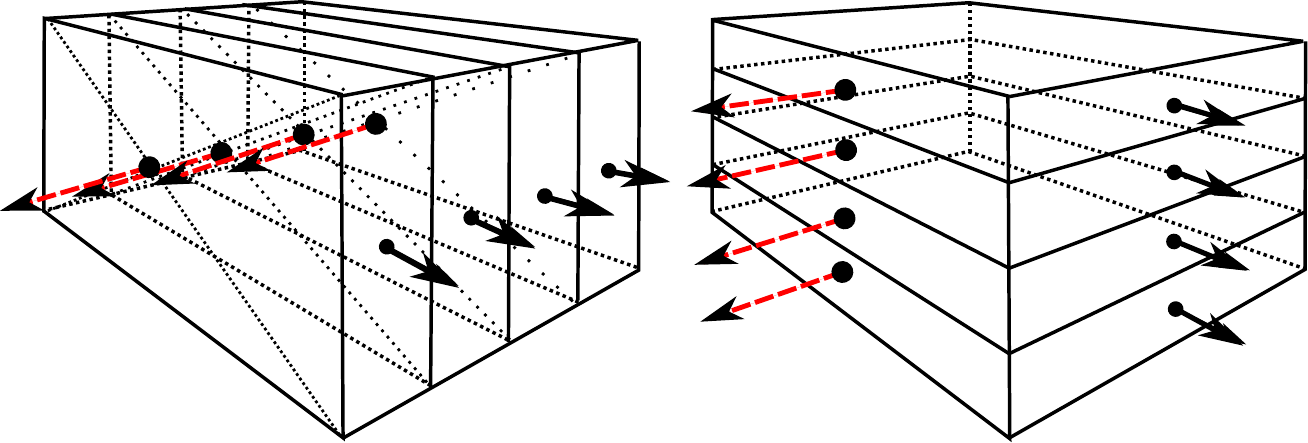}
                \caption{Case $2$:  $j\neq i$  in 
Eq. (\ref{eqn:XLES_CoupledAdvetionEquation}). Highly (coarsely) resolved 
advection in XLES-grid $3$ (left) (XLES-grid $2$ 
(right)).}
                \label{fig:case2}
        \end{subfigure}
	    \caption{XLES requires a coupled advection scheme. The advected 
velocity $\lspace{\vec{u}}{i}$ (bold arrows) is advected with the constant wave 
speed $c_{j}$ (\textcolor{red}{dashed} arrows) 
and is resolved in two XLES-grids. For the two relevant cases the discrete 
staggered grids within one 3D large scale cell (compare to figure 
\ref{fig:grids}) are shown in (a) and (b). 
}
	    \label{fig:Cases}
\end{figure}

Two situations occur (see figure \ref{fig:Cases}), as discussed here by a 
concrete example (generalization to other indices is trivial):
\begin{itemize}
 \item Case $1$: $j = i \stackrel{w.l.o.g.}{=}1$ in 
Eq. (\ref{eqn:XLES_CoupledAdvetionEquation}): 
Advection into $x_1$-direction; In both XLES-grids ${2}$ and ${3}$ the 
advection is coarsely resolved with $N_{\rm LES}$ grid cells.
 \item Case $2$: $2\stackrel{w.l.o.g.}{=}j\neq i\stackrel{w.l.o.g.}{=}1$ in 
Eq. (\ref{eqn:XLES_CoupledAdvetionEquation}): 
Advection into direction $x_{2}$; In XLES-grid $2$ the advection is 
resolved with $N_{\rm RSS}$ cells, respective with $N_{\rm LES}$ cells 
in XLES-grid ${3}$.
\end{itemize}

To spatially discretize the advection terms, a central difference method (CDM) 
on 
a 
staggered grid is used.

Two different time discretizations are deployed:
\begin{itemize}
 \item The highly resolved properties (using $N_{\rm RSS}$ cells) are 
discretized 
using a standard $2nd$ order 
Crank-Nicolson (CN) scheme in time (see \cite{CrankNicolson1947}).
We will refer to this fully discrete scheme as CN-CDM.
 \item The coarse grained resolved properties (using $N_{\rm LES}$ cells) are 
discretized using a $3$-stage $3rd$ order TVD 
Runge-Kutta (RK3) scheme (see \cite{Spiteri:2002}).
We will refer to this fully discrete scheme as RK3-CDM
\end{itemize}

The numerical properties of the numerical schemes without coupling are:
\begin{itemize}
 \item The advection scheme CN-CDM is stable, dissipation free and 
has low dispersion.
 \item The RK3-CDM scheme is found to be stable, producing little dispersion 
and 
dissipation.
\end{itemize}
These properties transfer to the coupled advection schemes (in Cases $1$ and 
$2$). 

Based on the expressions RK3-CDM and CN-CDM the coupled advection scheme in 
Case $1$ (Case $2$) is called RK3-RK3-CDM (CN-RK3-CDM).

To illustrate the coupling procedure w.l.o.g. for the CN-RK3-CDM scheme, we 
linearize and rewrite Eq. (\ref{eqn:XLES_CoupledAdvetionEquation}) in a 
semi-discrete form (in index notation). Hereby $\lspace{u}{3,1}$ is discretized 
in XLES-grid $3$ 
in Eq. (\ref{eqn:AdvectionExample3}) and $\lspace{u}{2,1}$ in XLES-grid $2$ 
in Eq. (\ref{eqn:AdvectionExample2}):
\begin{align}
   \label{eqn:AdvectionExample3}
   (\Delta t) \lspace{u}{3,1} =&
   \int_{RK3}
       c_2 \partial_{x_2}  {\lspace{u}{3,1}}
   \dint t    \\
	&+ [l_3^{-1}] 
	  [l_2]  
	  \left(
	    \int_{CN}
		c_2  \partial_{x_2}
		  \lspace{{u}}{2,1}
	    \dint t	 
	    -
	    \int_{RK3}
		c_2  \partial_{x_2}	
		  \lspaceall{{u}}{2,1}^{\rm LES}
	    \dint t
	   \right) \nonumber \\
	   \label{eqn:AdvectionExample2}
  (\Delta t) \lspace{u}{2,1} =&
   \int_{CN}
       c_2 \partial_{x_2}  {\lspace{u}{2,1}}
   \dint t    \\
	&+ [l_2^{-1}] 
	  [l_3]  
	  \left(
	    \int_{RK3}
		c_2  \partial_{x_2}
		  \lspace{{u}}{3,1}
	    \dint t	 
	    -
	    \int_{RK3}
		c_2  \partial_{x_2}	
		  \lspaceall{{u}}{3,1}^{\rm LES}
	    \dint t
	   \right) \nonumber.
\end{align}
Here the numerical time discretizations of the advection terms are indicated 
(CN and RK3). The discrete coupling operators $[l_k]$  and $[l_k^{-1}]$ (see 
Eq. (\ref{eqn:BoxFilter}) and the algorithm described 
in section \ref{ss:Refinement}) are 
discretized by an explicit Euler scheme in time.
Also the non-linear advection part which is not considered in the linear 
advection problem is discretized by an explicit Euler scheme in time.


 
The properties of the fully discrete linear coupled advection schemes are:
\begin{itemize}
 \item The coupled advection scheme converges to the analytical solution for 
$N_{\rm LES}\rightarrow 
\infty$ as investigated in \ref{app:Adv_highSSResolution} where the theoretical 
prediction e.g. by \cite{Tsai:2002} is reproduced. 
This property is 
required to numerically realize the LES limit and DNS limit of 
XLES (section \ref{s:XLES_LESlimit}).
 \item The coupled advection scheme including the deconvolution (see section 
\ref{ss:Refinement}) reproduces well defined large scale behavior, even by 
simultaneously transporting small scale properties, as 
investigated in
\ref{app:Adv_highSSResolution_SSProperty}.
\end{itemize}



The highly resolved CN-CDM scheme (using $N_{\rm RSS}$ cells) contributes to 
the coupled CN-RK3-CDM scheme (Case $2$) and additionally increases the 
numerical accuracy of the coupled scheme. This is neither required for a well 
defined and 
converging XLES scheme nor the LES limit (or DNS limit) of XLES. 
As investigated in \ref{app:Adv_highSSResolution} the highly resolved advection 
terms increase the coupled numerical accuracy up to a resolution ratio 
${N_{\rm 
RSS}/N_{\rm LES}} \lesssim 10$. With $N_{\rm RSS} \gtrsim 10 N_{\rm LES}$  
the overall numerical error is dominated by the 
coarsely resolved RK3-CDM scheme.


The coupled numerical schemes, simultaneously discretized by multiple 
XLES-grids, are found to be appropriate for linear advection. 
Important numerical properties, like stability and 
dissipative behavior, transfer from the well known underlying numerical 
discretizations to the coupled advection scheme.

There are other time schemes, that possibly can be adapted to the special 
requirements of the XLES time advancement, e.g.:
\begin{itemize}
 \item  An IMEX (implicit/explicit) time scheme (see e.g. 
	\cite{Cavaglieri:2015} and references cited therein) applied to XLES 
can in 
    principle 
	lead to a 
	high-order time integration for all terms (including non-linear 
advection     and 
	coupling), but to include the 
	coupling terms into such a high order time scheme requires identical 
	coefficients 
	for the implicit and explicit integration terms. 
 \item  Adapting split-explicit schemes (see e.g. \cite{Gadd:1978}) to the 
XLES coupled advection can especially decrease the dispersive effects 
that arise.
 \item  Large time step wave propagation schemes based on the work by 
\cite{LeVeque:1985} can perhaps improve the numerical properties of the applied 
CN scheme within the XLES advection scheme. 
\end{itemize}

For the sake of completeness the diffusion terms are discretized by a central 
difference scheme in space with a first order explicit (implicit) Euler scheme 
in the 
coarse (fine) resolved 
XLES-grid direction.

\subsection{XLES: Convergence of the XLES Approach}
\label{ss:Channel}

To verify the XLES model, we conduct a convergence study of \mbox{XLES-U} 
(unclosed 
XLES) for a fully developed turbulent flow, including the full diversity 
of non-linear advective effects: 
A turbulent channel flow with a friction Reynolds number 
$Re_{\tau}=395$ is computed.
DNS results by \cite{KAM99} (online available: \cite{Kawamura:2013}) 
are compared to the XLES-U outcome. 

The DNS is resolved with $N_{DNS}=192$ non-equidistant cells in the 
horizontal direction (between the walls). For the spatial discretization a 
second 
order central difference scheme is used. The time is discretized 
using a Crank-Nicolson scheme 
for the wall-normal non-linear terms and a second order Adams-Bashforth 
scheme for other terms (results for higher resolutions and higher order 
schemes are also available online). This numerical scheme is comparable to the 
XLES-U 
numerical scheme (see section \ref{ss:XLES_Numerics}).

For the XLES-U convergence study, the number of equidistant 3D large scale 
cells $N_{\rm LES}$ is increased using the values 
$N_{\rm LES}=\{16,32,64\}$ while  
 $N_{\rm RSS}=512$ is kept constant.
 
 The time step size is limited by:
\begin{align}
  \Delta t = CFL\, \min_{i,k}\left( \frac{\Delta x_{k,i}^{\rm RSS} }{ 
\lspace{u}{k,i}} \right) ,\, i,k=\{1,2,3\}
\end{align} 
 with a Courant-Friedrichs-Lewy number 
$CFL=0.45$ and the small scale cell size $\Delta x_{k,i}^{\rm RSS}$.

To produce statistically significant results, the flow is averaged for 
$t_{ave} \geq 25$ non-dimensional time units (compared to $t_{ave}=20$ for DNS) 
after reaching a statistically steady state.

Additionally equidistant discretized LES-U (LES limit of XLES-U with $N_{\rm 
RSS} = N_{\rm LES}$, see section 
\ref{s:XLES_LESlimit}) channel flow results are compared to XLES-U.
The latter resolves selected advective and diffusive small scale effects even 
without using a SGM. Those can be identified by comparing the XLES-U and 
LES-U channel flow results.

\begin{figure}
        \centering      
        \begin{subfigure}[b]{0.49\textwidth}
	      \centering               
                \includegraphics[width=0.9\textwidth]{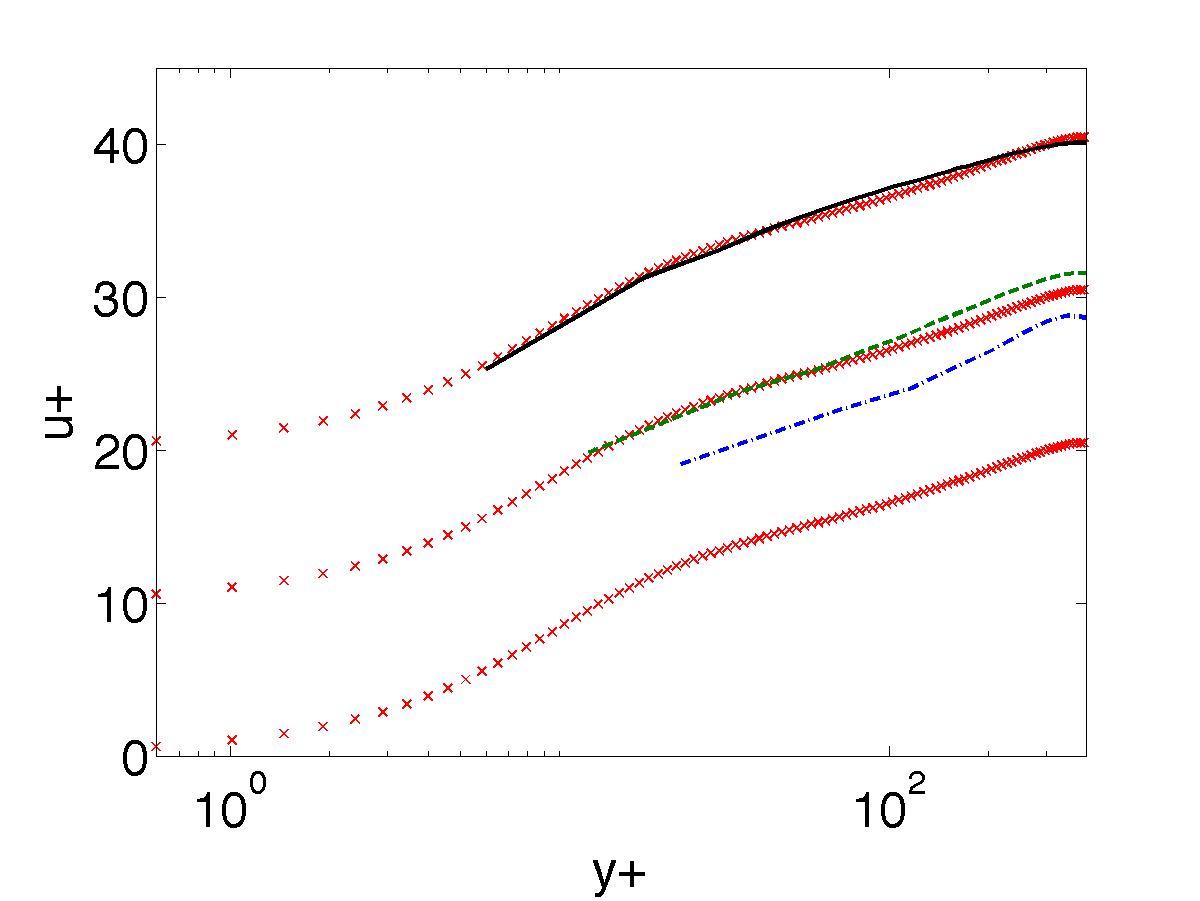}
                \caption{ LES: law of the wall ($\lspaceall{u}{1}^{\rm LES}$). 
Profiles shifted with increasing $N^{\rm LES}$.}
                \label{fig:umLES}
        \end{subfigure}
        \begin{subfigure}[b]{0.49\textwidth}
	      \centering        
                \includegraphics[width=0.9\textwidth]{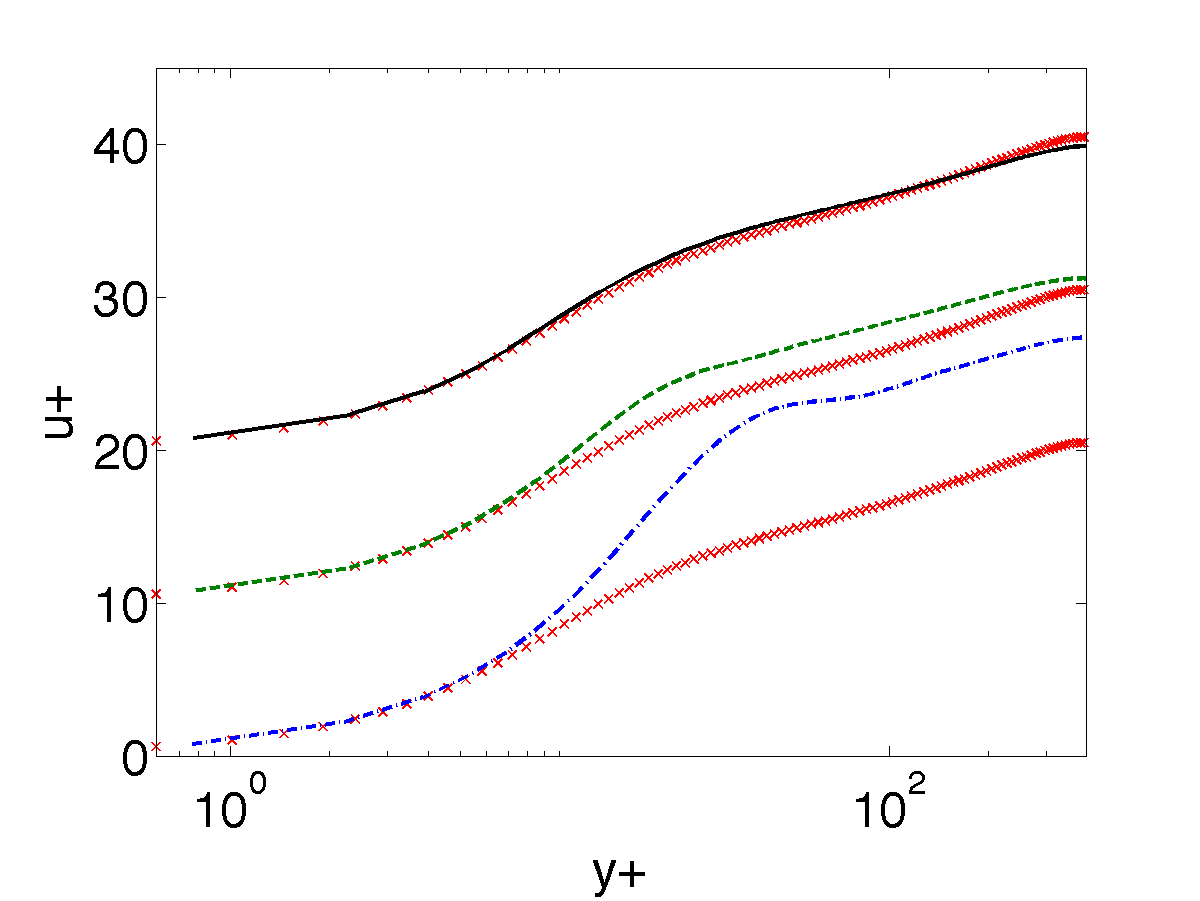}
                \caption{ XLES: law of the wall ($\lspace{u}{2,1}$). Profiles 
shifted with increasing $N^{\rm LES}$.}
                \label{fig:umXLES}
        \end{subfigure}          
        \\
        \begin{subfigure}[b]{0.49\textwidth}
	      \centering               
                \includegraphics[width=0.9\textwidth]{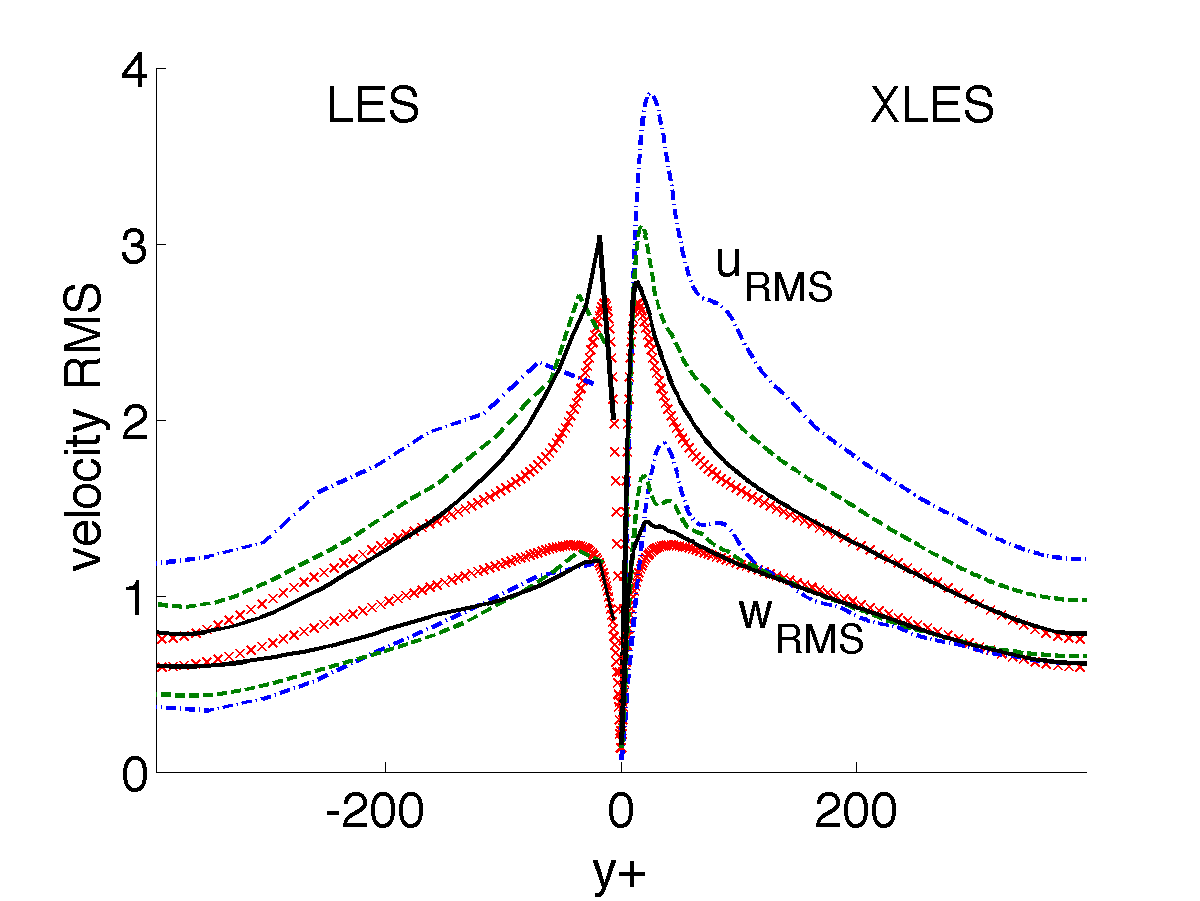}
                \caption{ Streamwise ($u_{\rm RMS}$) and spanwise  
($w_{\rm RMS}$) velocity RMS.}
                \label{fig:uvwR}
        \end{subfigure}
        \begin{subfigure}[b]{0.49\textwidth}
	      \centering               
                \includegraphics[width=0.9\textwidth]{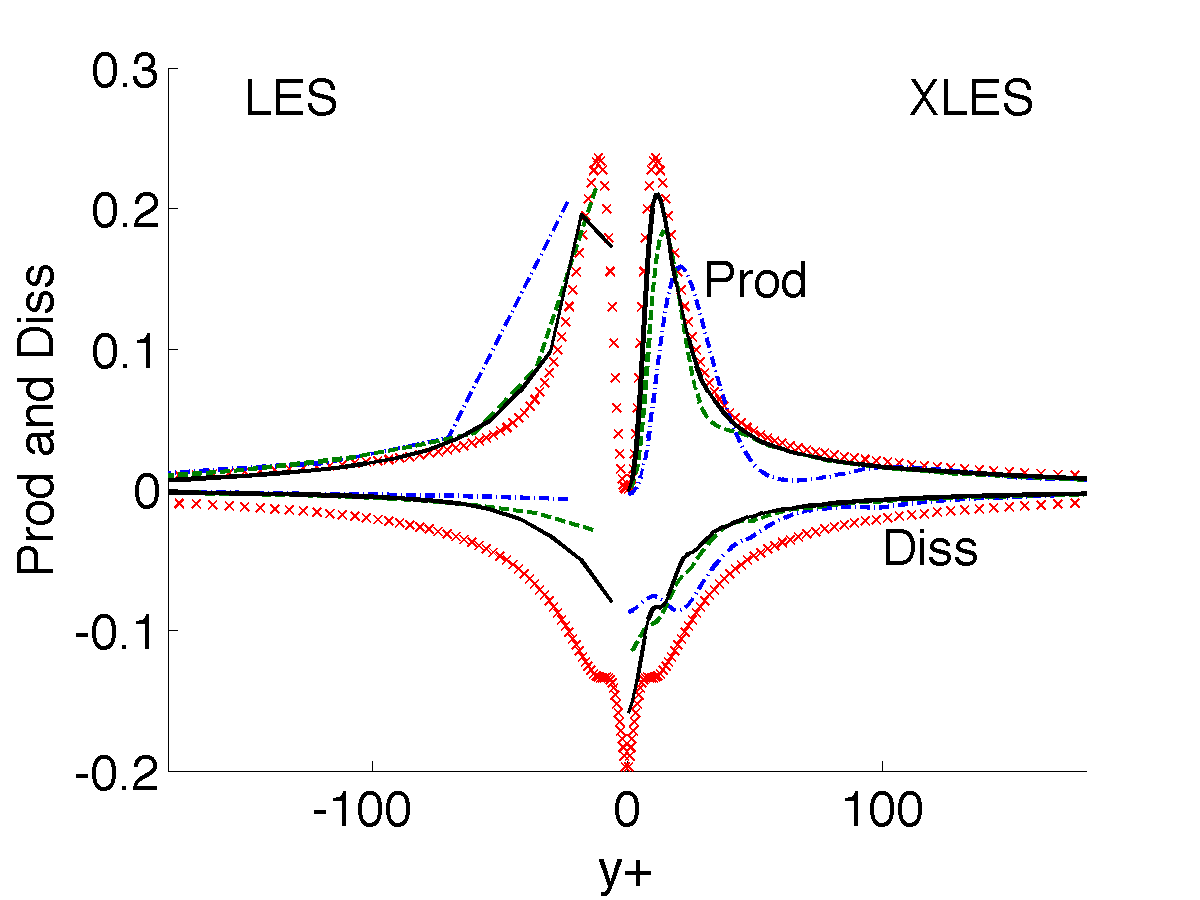}
                \caption{Production (Prod) and Dissipation (Diss) of the 
turbulent kinetic energy.}
                \label{fig:ProdDiss}
        \end{subfigure}\\
        \begin{subfigure}[b]{0.49\textwidth}
	      \centering               
                \includegraphics[width=0.9\textwidth]{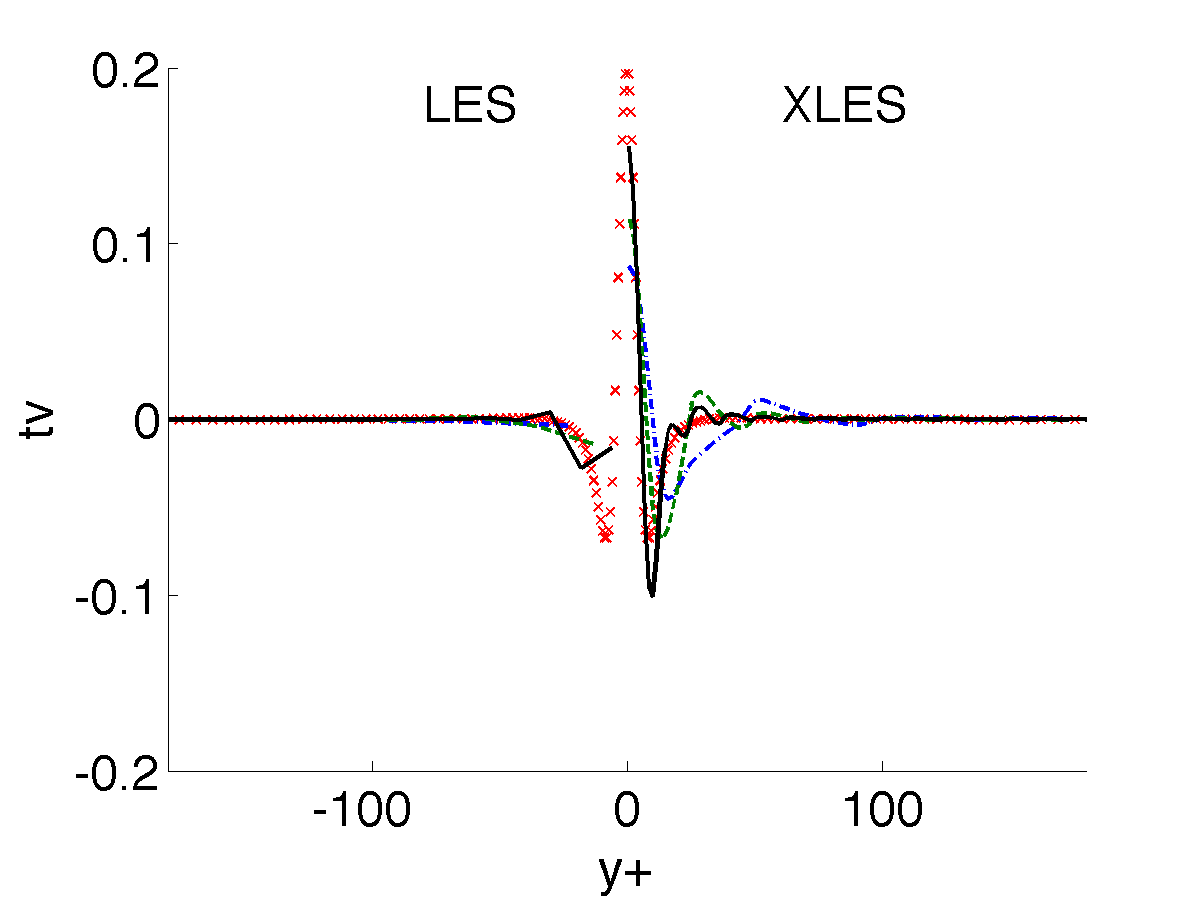}
                \caption{Viscous transport of the turbulent kinetic 
energy (tv).}
                \label{fig:Tv}
        \end{subfigure}
        \begin{subfigure}[b]{0.49\textwidth}
	      \centering               
                \includegraphics[width=0.9\textwidth]{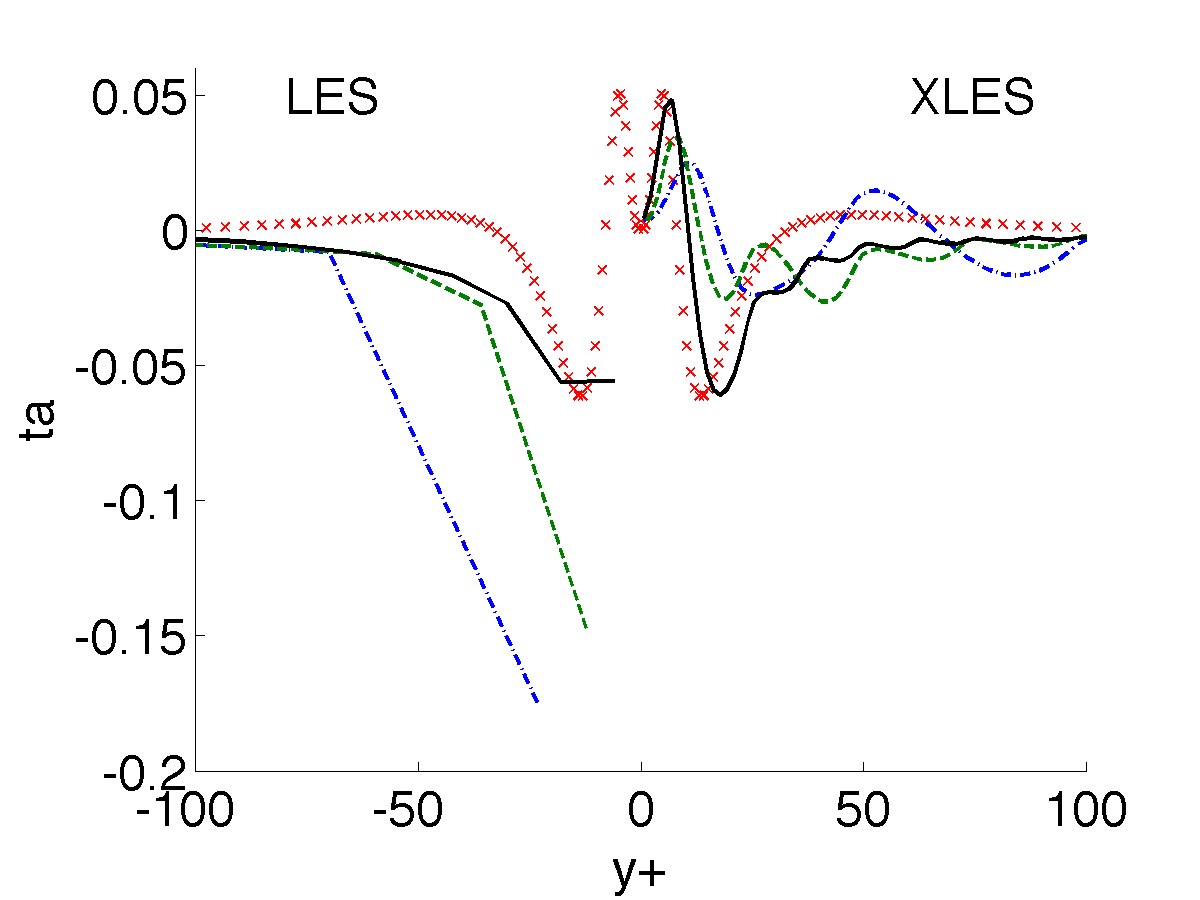}
                \caption{Advective transport of the turbulent kinetic 
energy (ta).}
                \label{fig:Ta}
        \end{subfigure}        
	    \caption{Turbulent channel flow results for DNS 
(\textcolor{red}{small crosses}), LES-U, and XLES-U with $N_{\rm 
LES}=16$ 
(\textcolor{blue}{dash-dotted}), $N_{\rm LES}=32$ 
(\textcolor{darkgreen}{dashed}), and $N_{\rm LES}=64$ 
(\textcolor{black}{solid}). The XLES-U small scales are resolved using 
$N_{\rm RSS}=512$ cells. 
XLES-U mean and statistical flow properties are based on the 
velocity field perpendicular 
to the channel walls $\lspace{u}{2,i}$ 
(in the LES limit: $\lspace{u}{2,i}= \lspaceall{u}{i}^{\rm LES}$).}
	    \label{fig:Channel_Um} 
\end{figure}

Figure \ref{fig:Channel_Um} illustrates the results:
The mean velocity profiles computed by XLES-U (see figure \ref{fig:umXLES}) and 
LES-U (see figure \ref{fig:umLES}) are compared to DNS.


Additionally the streamwise and spanwise velocity RMSs (see figure 
\ref{fig:uvwR}) and the budget terms of the turbulent kinetic energy (see 
figure \ref{fig:ProdDiss}- \ref{fig:Ta}) are shown for XLES-U, LES-U, and DNS.

Both XLES-U and LES-U show convergence towards the DNS results with 
increasing 3D resolution.
Following arguments in section \ref{s:XLES_LESlimit} 
this implies the XLES error term $\Restspace{XLES}$ to be small. 
Additionally the XLES non-linear terms 
(and its LES limit) are sufficiently represented and converging, even if the 
under-resolved velocity profile (figure 
\ref{fig:umLES} and \ref{fig:umXLES}) indicates noticeable numerical 
dissipation.

XLES-U is able to accurately represent diffusive effects, e.g. the laminar 
sublayer near the walls, independent of the 3D large scale resolution and 
reproduces basic advective effects including turbulence with $N_{\rm 
LES} \gtrsim 32$ 3D cells for $Re_{\tau}=395$.
Additionally the turbulence statistics profit from the additional small 
scale effects 
represented by XLES compared to LES-U even with low 3D resolutions.

In part II (\cite{Glawe:2014B}) ODT is demonstrated to represent the turbulent 
effects not resolved by XLES-U, even with very coarse 3D resolutions (e.g. 
$N_{\rm LES}=16$).

\section{Conclusions}
\label{s:Conclusions}

In this work XLES, an extended LES model, is introduced. 
This approach is intended as a basis for a new class of turbulence 
models, tailored to describe a (XLES specific) macrostructure of highly 
turbulent flows 
including fully resolved molecular diffusion in domains of moderate complexity.

An innovative XLES filter strategy is derived, including a numerical 
representation of the resulting 2D filtered XLES equations by three coupled 
XLES-grids.
This coupling requires a high order deconvolution function for 1D box 
filtered fields, wherefore a stabilized version of the algorithm introduced by 
\cite{RC-Schmidt2010} is applied. 

The 3D filtered LES-U (unclosed LES) equations and DNS are distinguished 
limits of 
XLES-U (unclosed XLES).

We verify the XLES-U equations:
\begin{itemize}
 \item By analyzing the coupled (linear) advection scheme occurring in XLES.
 \item By performing a convergence study: The
turbulent channel flow with $Re_{\tau}=395$ is compared to DNS and LES-U.
Compared to LES-U low 3D resolutions are required in XLES-U to reproduce 
fundamental flow statistics, especially in the near wall region. E.g. the 
laminar 
sublayer is 
fully represented independently of the 3D resolution. 
\end{itemize}

An XLES closure can be provided by the ODT model, which is able to capture the 
full turbulent cascade of highly 
turbulent flows, as introduced in part II 
(\cite{Glawe:2014B}). Recently \cite{Meiselbach2015} introduces channel flow 
results up to $Re_{\tau} = 600000$ using adaptive ODT by \cite{Lignell2012}.
This ODTLES model enables the representation of highly turbulent flows in 
domains of moderate complexity with coarse 3D resolutions basically independent 
of turbulent scales and at once represents diffusive and (turbulent) 
advective effects down to the 
molecular level using ODT. Hereby the 3D resolution needs to be sufficient to 
resolve the domain and possible secondary effects. E.g. due to secondary 
instabilities within a duct flow, the 3D resolution indirectly depends on the 
Reynolds number (investigated in  part II by \cite{Glawe:2014B}). 
The XLES model is especially advantageous for turbulent flows with crucial 
small scale effects, e.g. 
with important Prandtl or Schmidt number effects, buoyant stratification 
effects, or in the field of combustion.


In XLES a potential bottleneck in highly 
resolved incompressible simulations, the pressure handling, is only computed on 
a very coarse 3D grid and small scale pressure effects might be modeled.
Closed XLES (e.g. ODTLES) is a promising and highly 
parallelizable approach to investigate fundamental atmospheric turbulent
flows. 





\section*{Acknowledgments}
\label{s:Acknowledgements}

The authors would like to thank H. Kawamura and colleagues for providing 
DNS results online (\cite{Kawamura:2013}).
This work was 
supported by the Brandenburg University of 
Technology Cottbus-Senftenberg, the Helmholtz graduate research school GeoSim, 
and the Freie Universit\"at Berlin.

\appendix

\section{XLES: Vector Notation}
\label{app:XLES_MatrixDerivation}

We introduce an alternative to the vector notation in section 
\ref{s:XLES} by writing the XLES velocity scales in index notation. 
The XLES vector $\lspace{\vec{u}}{i}$ includes the XLES vector 
elements $\lspace{{u}}{k,i}$ represented in XLES-grids $k$, $k=\{1,2,3\}$. 

In index notation the resolved small scale (RSS) terms $s_1 l_2 l_3$, $l_1 s_2 
l_3$, and $l_1 l_2 s_3$ can be expressed by one term: $s_k l_{k\oplus1} 
l_{l\oplus2}$, 
where $\{k,k\oplus1,k\oplus2\}$ is a positive permutation of $\{1,2,3\}$ 
(the 1D 
filter operators are commutable).
The operator $\oplus$ denotes the positive permutation:
\begin{align}
 \label{eqn:IndexDef}
  q \oplus x = ((q+x-1) \mod 3)+2
\end{align}
for $q=\{1,2,3\}$ and $x=\{1,2\}$ (valid in three dimensions). 

The XLES resolved velocity scales (Eq. (\ref{eqn:XLES_AllResolvedScales})) are 
written in index notation (using the 
operator $\oplus$):
\begin{align}
  \lspaceall{u}{i} &= \left[  {  l_1 l_2 l_3 +s_1 l_2 l_3 + l_1 s_2 l_3 + 
l_1 l_2 s_3} \right] u_i  \nonumber \\
        =& \left[   l_1 l_2 l_3 + 
\sum_{q=1}^3 s_q l_{q\oplus1} l_{q\oplus2} 
\right] u_i  \label{eqn:XLESgridAnsatz} \\
        =& \left[  \sum_{q=1}^3 \left( l_q l_{q\oplus1} l_{q\oplus2} +  s_q 
l_{q\oplus1} l_{q\oplus2} \right)  
 - 2 l_1 l_2 l_3  \right] u_i   \nonumber.
\end{align}

The 3D large scale operator $[l_1 l_2 l_3]$ in the 
last row of Eq. (\ref{eqn:XLESgridAnsatz}) can be expressed 
in terms of $[l_q l_{q\oplus1} l_{q\oplus2} +  s_q l_{q\oplus1} l_{q\oplus2}]$:
\begin{align}
  - 2 l_1 l_2 l_3 =& - \sum_{q=1}^3 (1-\delta_{qk} ) (l_q l_{q\oplus1} 
l_{q\oplus2}) \nonumber \\
 =& - \sum_{q=1}^3 (1-\delta_{qk} ) ( (l_q l_q +l_q- l_q l_q) l_{q\oplus1} 
l_{q\oplus2} ) \label{eqn:App_3DLargeScale} \\ 
 =& - \sum_{q=1}^3 (1-\delta_{qk} ) l_q ( l_q l_{q\oplus1} 
l_{q\oplus2} +  \underbrace{(1-l_q)}_{=s_q} l_{q\oplus1} 
l_{q\oplus2})  \nonumber
\end{align}
with the Kronecker delta operator defined in Eq. 
(\ref{eqb:DiracDeltaOperatorMatrix}).

Here an additional arbitrary index $k=\{1,2,3\}$ is introduced. 
Eq. (\ref{eqn:App_3DLargeScale}) is satisfied for $k=1$, $k=2$, and $k=3$.
This index $k$ spans the XLES vector. 

Insertion of the 3D large scale operator (Eq. (\ref{eqn:App_3DLargeScale})) 
into 
the XLES resolved velocity scales (Eq. (\ref{eqn:XLESgridAnsatz})) leads to:
\begin{align}
  \label{eqn:XLESgridAnsatz2}
  \lspaceall{u}{k,i} &=  \left[  \sum_{q=1}^3 (1 - (l_q - l_q \delta_{qk})) 
\left( l_q 
l_{q\oplus1} l_{q\oplus2} +  s_q 
l_{q\oplus1} l_{q\oplus2} \right) \right] u_i  , \; 
k=\{1,2,3\} .
\end{align}
This XLES resolved velocity $\lspaceall{u}{k,i}$ reproduces exactly the XLES 
resolved velocity scales in Eq. (\ref{eqn:XLES_AllResolvedScales}) including 
the XLES coupling  $\mat{C}\, \mat{s}^{1D} \mat{l}^{2D}$.
XLES is interpreted as approach filtering the Navier-Stokes equations 
by applying the operator $ \left[  {  l_1 l_2 l_3 +s_1 l_2 l_3 + l_1 s_2 l_3 
+l_1 l_2 s_3} \right]$ compared to $ \left[  {  l_1 l_2 l_3 } \right]$ in LES 
(see table \ref{tab:CompareModels}).

\section{XLES Mass Conservation: Resolved Small Scales}
\label{app:Mass_ResolvedSmallScales}

Divergence-free  XLES velocity fields $\lspace{\vec{u}}{i}$ are guaranteed, if 
the 3D large scale velocity field $\lspaceall{\vec{u}}{i}^{LES}= 
[l_k]\lspace{u}{k,i}$ fulfills
three conditions (see section \ref{s:mass}):
\begin{itemize}
 \item Con.$1$: 
 $\lspaceall{u}{k,i}^{LES}$ is consistent 
($\lspaceall{u}{i}^{LES}=\lspaceall{u}{k,i}^{LES}$)
 \item Con.$2$:
  $\lspaceall{u}{i}^{LES}$ is divergence free, enforced by 3D standard approach.
 \item Con.$3$:
 $[l_k]$ is a discrete 1D box filter (defined in Eq. (\ref{eqn:BoxFilter})) in 
$x_k$-direction.
\end{itemize}

The XLES velocity fields $\lspace{\vec{u}}{i}$ are divergence free as proved 
within one 3D large scale cell of the size $\Delta {x_k}$ in $x_k$-direction 
(operator $\oplus$ is defined in Eq. (\ref{eqn:IndexDef})):
  \begin{align}
  0 &=\int_{-\frac{\Delta x_k}{2}}^{\frac{\Delta x_k}{2}} \sum_{j=1}^3 
\partial_{x_j} (\lspaceall{u}{k,j}^{\rm LES} + \rsspace{u}{k,j})
\dint x_k
  = \int_{-\frac{\Delta x_k}{2}}^{\frac{\Delta x_k}{2}} \sum_{j=1}^3 
\partial_{x_j} \lspace{u}{k,j} \dint x_k
   \\
  &= \int_{-\frac{\Delta x_k}{2}}^{\frac{\Delta x_k}{2}} 
  \partial_{x_k} \lspace{{u}}{k,k} \dint{x_k}
  + \int_{-\frac{\Delta x_k}{2}}^{\frac{\Delta x_k}{2}} 
  \partial_{x_{k \oplus 1}} \lspace{u}{k,{k \oplus 1}} \dint{x_k}
  + \int_{-\frac{\Delta x_k}{2}}^{\frac{\Delta x_k}{2}} 
  \partial_{x_{k \oplus 2}} \lspace{{u}}{k,{k \oplus 2}} \dint{x_k}\nonumber\\
  &= \lspace{u}{k,k}\left(\frac{\Delta x_k}{2}\right) 
  - \lspace{u}{k,k}\left(-\frac{\Delta x_k}{2}\right)
  +\partial_{x_{k \oplus 1}} \int_{-\frac{\Delta x_k}{2}}^{\frac{\Delta 
x_k}{2}} 
   \lspace{{u}}{k,{k \oplus 1}} \dint{x_k}
  + \partial_{x_{k \oplus 2}} \int_{-\frac{\Delta x_k}{2}}^{\frac{\Delta 
x_k}{2}} 
  \lspace{{u}}{k,{k \oplus 2}} \dint{x_k} \nonumber\\
  &\stackrel{{\rm Con.}3}{=}
    \lspaceall{u}{k,k}^{\rm LES} \left(\frac{\Delta x_k}{2}\right) 
  - \lspaceall{u}{k,k}^{\rm LES} \left(-\frac{\Delta x_k}{2}\right)
  +\partial_{x_{k \oplus 1}} \lspaceall{u}{k,k \oplus 1}^{\rm LES}
  + \partial_{x_{k \oplus 2}} \lspaceall{u}{k,k \oplus 2}^{\rm LES} \nonumber \\
  &= \int_{-\frac{\Delta x_k}{2}}^{\frac{\Delta x_k}{2}} \sum_{j=1}^3 
\partial_{x_j} \lspaceall{u}{k,j}^{\rm LES} \dint x_k 
   \stackrel{{\rm Con.}1}{=} \int_{-\frac{\Delta 
x_k}{2}}^{\frac{\Delta x_k}{2}} \sum_{j=1}^3 
   \partial_{x_j} \lspaceall{u}{j}^{\rm LES} \dint x_k  
   \stackrel{{\rm Con.}2}{=} 0
\nonumber.
\end{align}
for each XLES-grid $k=\{1,2,3\}$. 
Here a staggered grid is used, leading to
$\lspace{u}{k,k}\left(\frac{\Delta 
x_k}{2}\right) = \lspaceall{u}{k,k}^{\rm LES} \left(\frac{\Delta 
x_k}{2}\right)$ without additional interpolation.

This in particular means the resolved small scale (RSS) velocity fields 
$\rsspace{\vec{u}}{j}$ are divergence free by construction. 

\section{XLES: Numerical Advection Scheme}
\label{app:LinearAdvection}

The XLES equations are discretized using multiple, coupled XLES-grids. 
This discretization includes a coupled advection scheme, 
possessing uncertain numerical properties.

The coupled advection scheme is investigated and verified by computing a linear 
wave propagation.
The initial condition
\begin{align}
  \label{eqn:Sin}
  {u}_{1} = \sin(4 \pi x_1)
\end{align}
and periodic boundary conditions are used.
The number of cells for the resolved small scale fields $N_{\rm RSS}$ and the 
3D 
large scale field
$N_{\rm LES}$ are varied.
The wave is propagated for 10 wavelengths ($t= 5 
\,L/c_j$; $L\equiv1$; constant wave speed $c_j$).

All calculations are done with a Courant-Friedrichs-Lewy number $CFL=0.45$ 
limiting the time step size 
\begin{align}
  \Delta t = CFL\, \min_{i,k}\left( \frac{\Delta x_{k,i}^{\rm LES} }{ 
\lspace{u}{k,i}} \right) ,\, i,k=\{1,2,3\}.
\end{align}

Please note the non-linear problem in section \ref{ss:Channel} is limited to 
time steps based on $\Delta x_{k,i}^{\rm RSS}$ instead of $\Delta x_{k,i}^{\rm 
LES}$, which is investigated in part II 
(\cite{Glawe:2014B}) in detail. 

\subsection{Coupled Advection Schemes: RK3-RK3-CDM and CN-RK3-CDM}
\label{app:RK3-CDM}

\cite{Zha:2003} showed that RK2-CDM (central difference in space and 2nd 
order Runge Kutta in time) is 
unstable, while 
RK4-CDM (4th order Runge Kutta in time) is stable and fully dissipation free.

\begin{figure}
        \centering
	    \includegraphics[width=0.65\textwidth]{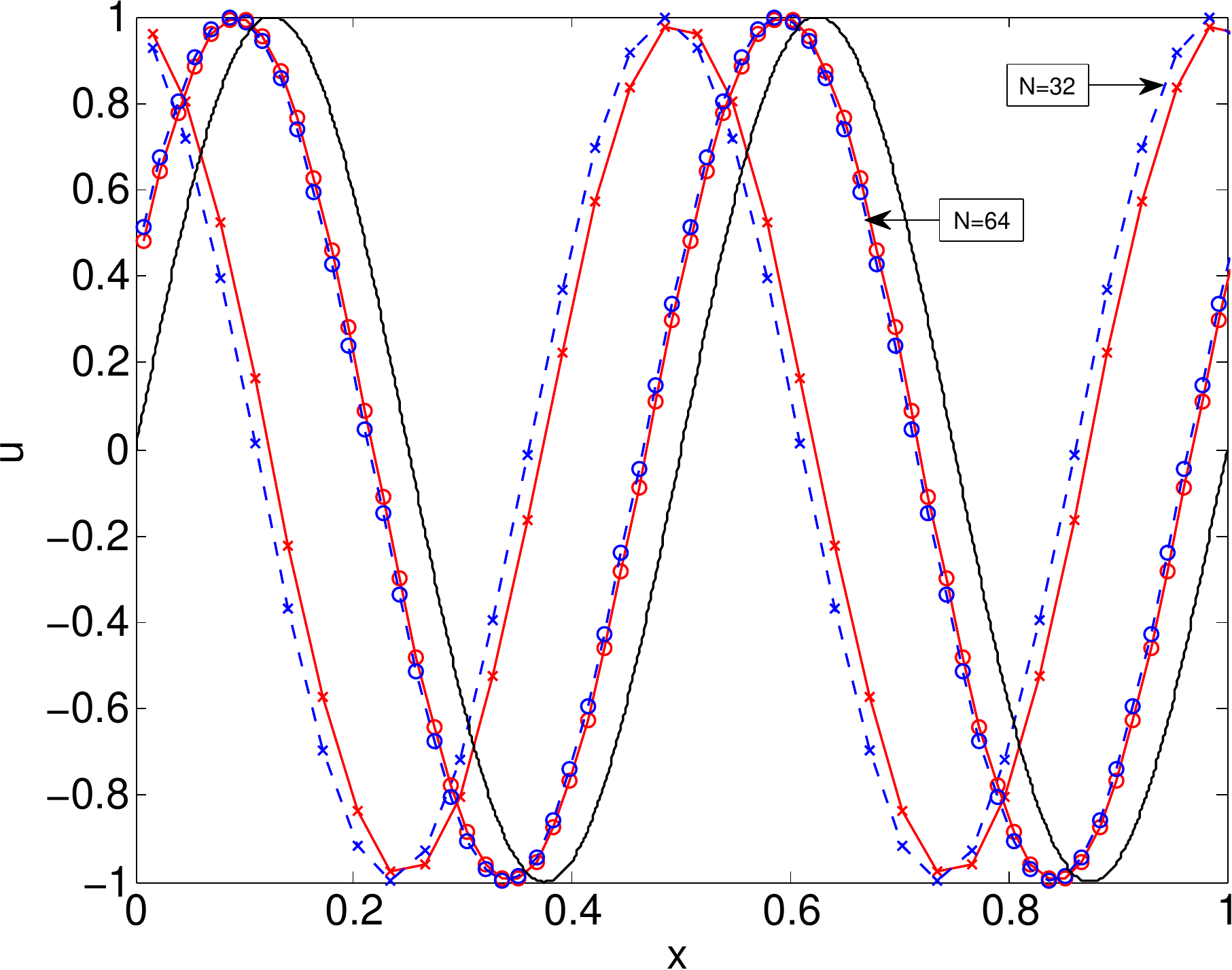}
	    \caption{A linear wave is propagated with constant velocity 
$\lspace{\vec{u}}{j} = \vec{c}_j$  over 10 wavelengths.
The coupled numerical schemes CN-RK3-CDM (\textcolor{blue}{dashed}) and 
RK3-RK3-CDM (\textcolor{red}{solid}) with 
$N=N_{\rm RSS}=N_{{\rm LES}}=\{32,64\}$ are shown. The black line is the 
analytical result. }
	    \label{fig:CNvsRK}
\end{figure}

We found the coupled advection schemes RK3-RK3-CDM (Case $1$ in section 
\ref{ss:XLES_Numerics}) and CN-RK3-CDM (Case $2$ in section 
\ref{ss:XLES_Numerics}) to be stable 
and slightly dispersive (see figure 
\ref{fig:CNvsRK}). Additionally RK3-RK3-CDM is little dissipative. 
The difference between RK3-RK3-CDM and CN-RK3-CDM is not significant.
With increasing 3D resolution ($N_{\rm LES}$) the dispersive behavior 
decreases (convergence). 

\subsection{Coupled Advection Scheme: High Small Scale Resolution}
\label{app:Adv_highSSResolution}

Numerical convergence for the XLES coupled advection scheme is 
unconditionally obtained for increasing 3D resolution ($N_{\rm 
LES}\rightarrow \infty$). This is valid for both coupled schemes: RK3-RK3-CDM 
and CN-RK3-CDM.

The convergence for the CN-RK3-CDM scheme with increasing small scale 
resolution 
($N_{\rm RSS}\rightarrow \infty$) and two choices of constant 3D 
resolution $N_{\rm LES} = \{16,64\}$ is additionally investigated.
This kind of numerical convergence is not required to ensure a well defined and 
converging XLES approach.

\begin{figure}
        \centering
	    \includegraphics[width=0.65\textwidth]{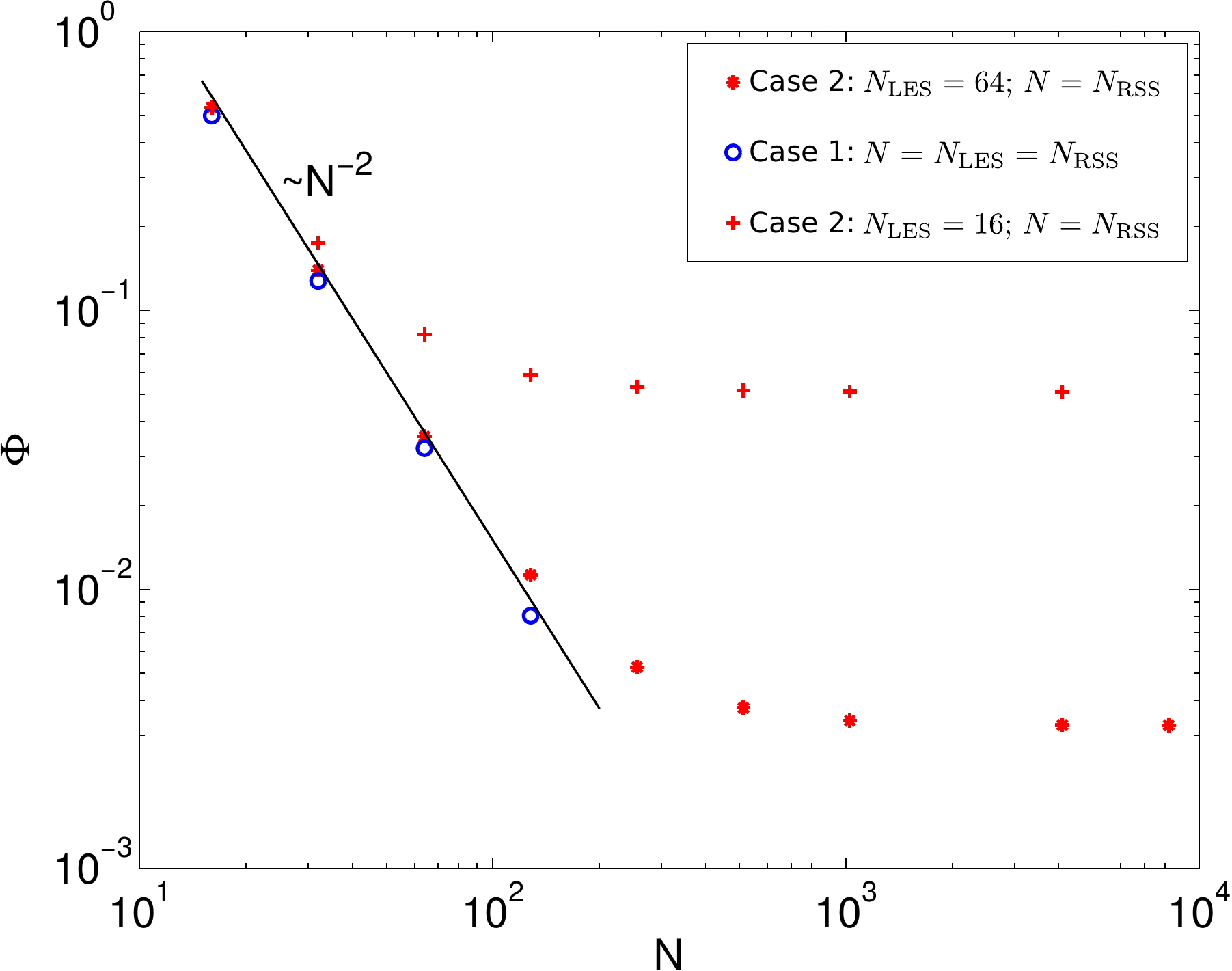}
	    \caption{$\Phi$ is the phase difference between the analytical 
solution and the numerical result.
Case $1$: RK3-RK3-CDM scheme with $N=N_{\rm LES}=N_{\rm RSS}$.
Case $2$: CN-RK3-CDM  scheme with $N=N_{\rm RSS}$ and $N_{\rm 
LES}=\{16, 64\}$. 
For a CN-CDM scheme \cite{Tsai:2002} analytically predict the behavior 
: $ \Phi \sim N^{-2}$.} 
	    \label{fig:Phase} 
\end{figure}

A phase shift $\Phi$ describes the numerical dispersion and thus measures the 
numerical accuracy (see figure \ref{fig:Phase}).

We find the numerical accuracy of the coupled advection scheme 
increasing up to a resolution ratio ${N_{\rm RSS} / N_{\rm LES}} \lesssim 10$.
With higher ratios the numerical error of the coarse resolved scheme (RK3-CDM) 
dominates the coupled numerical error.

Thus the numerical accuracy is increased by increasing the resolved small 
scales within 
XLES up to some limit in addition to possibly resolving additional physical 
effects.

\subsection{Coupled Advection Scheme: Spectral Scale Separation}
\label{app:Adv_highSSResolution_SSProperty}

In this section the CN-RK3-CDM scheme demonstrates the preservation of 
(spectrally separated) large and small scale properties which are represented 
w.l.o.g. in 
XLES-grid $2$.

A multi-scale wave with the initial condition
\begin{align}
\label{eqn:SinMultiScale}
 {u}_{{2},1} &= \sin(4 \pi x_{{2},1})  + 0.2 \,\sin(128 \pi 
x_{{2},1}) \\
 {u}_{{3},1} &= [l_{{3}}^{-1}] [l_{2}] 
u_{{2},1} = \sin(4 
\pi x_{{3},1})
\end{align}
is propagated for $10$ wavelengths using $N_{\rm LES}=64$ and $N_{\rm 
RSS}=4096$ cells in the $x_2$-direction (advanced by Eq. 
(\ref{eqn:AdvectionExample2}) and (\ref{eqn:AdvectionExample3})). 

Note that the higher mode $0.2 \,\sin(128 \pi 
x_{{2},1})$ is only resolved within XLES-grid $2$ while XLES-grid $3$ only 
captures the lower mode. 

\begin{figure}
        \centering
	  \includegraphics[width=0.65\textwidth]{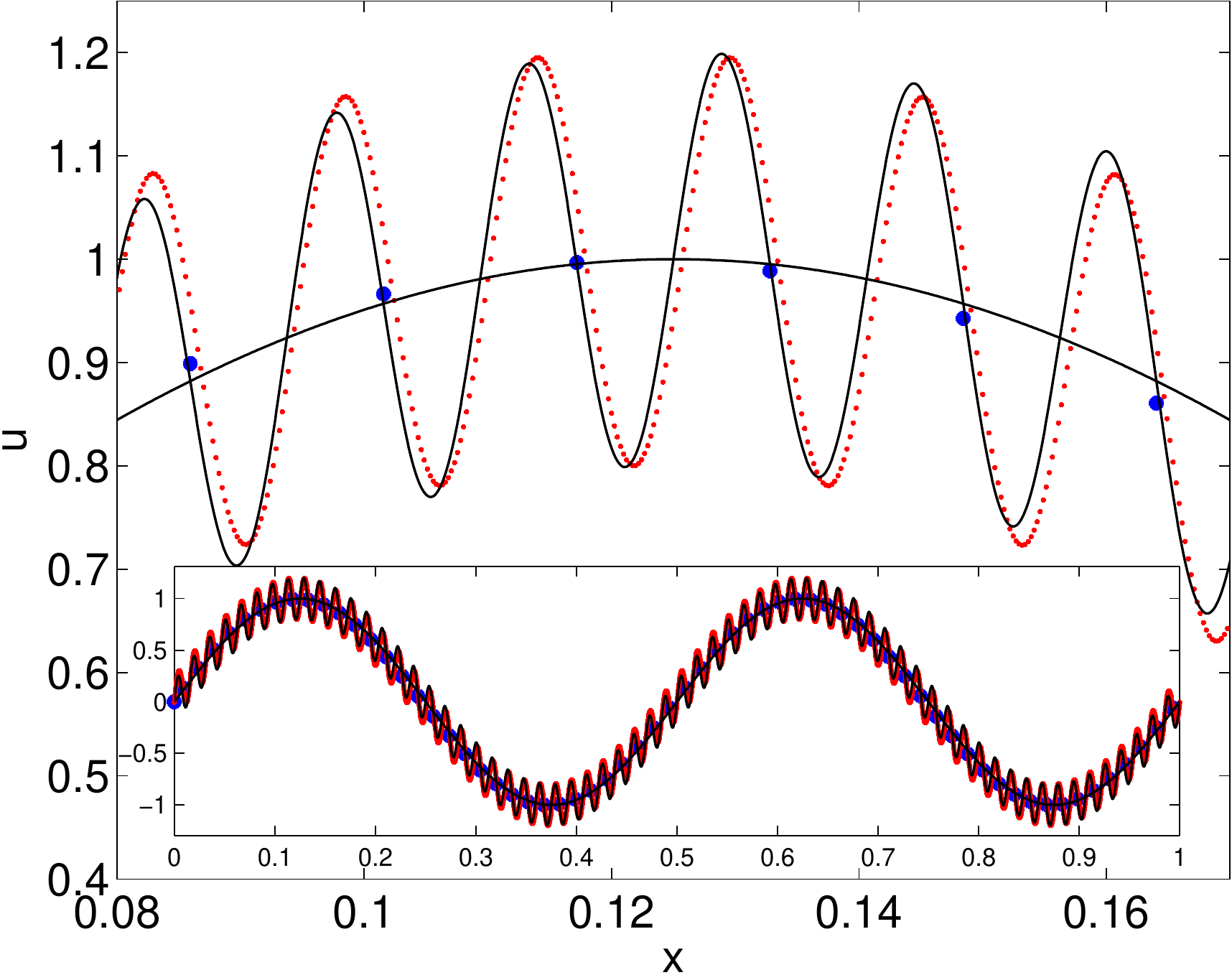}
	  \caption{
A multi-scale wave with the initial condition in Eq. (\ref{eqn:SinMultiScale}) 
(lines) propagates for $10$ wavelengths.  
The full domain (small box) and one wave 
peak are shown. 
The \textcolor{blue}{big points} corresponds to the 3D large scale field
($N_{\rm LES}=64$ cells). 
The \textcolor{red}{small points} corresponds to the resolved small scales
($N_{\rm RSS}=4096$). The black line corresponds to the analytical result.}
      \label{fig:URall}
\end{figure}

The box filtered and small scale properties are preserved, but little dispersed 
due to numerical 
effects (see figure \ref{fig:URall}). 

 \bibliographystyle{elsarticle-harv} 
 \bibliography{glawe.bib}


%
%
%
\end{document}